\documentclass[entropy,article,submit,pdftex]{mdpi}
\usepackage[T1]{fontenc}
\usepackage[utf8]{inputenc}
\usepackage{microtype}
\usepackage{amsmath,amssymb}
\usepackage{graphicx}
\usepackage{flafter}    
\usepackage{xcolor}

\providecommand{\methodname}{the cross-entropy audit (CEA)}

\newcommand{\SMentropy}{Supplementary Materials~S9}
\newcommand{\SMdeclarations}{Supplementary Materials~S13}
\newcommand{\SMdesign}{Supplementary Materials~S1}
\newcommand{\SMaccounts}{Supplementary Materials~S7}
\newcommand{\SMhuman}{Supplementary Materials~S2}
\newcommand{\SMaudit}{Supplementary Materials~S3}
\newcommand{\SMjitter}{Supplementary Materials~S10}
\newcommand{\SMrepair}{Supplementary Materials~S11}
\newcommand{\SMbounded}{Supplementary Materials~S14}
\newcommand{\SMomega}{Supplementary Materials~S4}
\newcommand{\SMblind}{Supplementary Materials~S5}
\newcommand{\SMpilot}{Supplementary Materials~S6}
\newcommand{\SMalttests}{Supplementary Materials~S15}
\newcommand{\SMarchive}{Supplementary Materials~S12}

\Title{A Cross-Entropy Audit (CEA) for Consciousness Tests Moved from Brain to Large Language Model}
\Author{Pieter van Rooyen $^{1,*}$}
\AuthorNames{Pieter van Rooyen}
\address{%
$^{1}$ Stellenbosch University, Department of Electrical and Electronic Engineering, Bosman St, Stellenbosch Central, Stellenbosch, 7600, South Africa; pgwvanrooyen@sun.ac.za; ORCID:\ 0009-0005-7708-8236}
\corres{Correspondence: pgwvanrooyen@sun.ac.za}

\conference{}
\firstnote{This paper is a contribution to the Entropy 2026 special issue on the subjects of the conference (Barcelona, Spain, 1--3 July 2026). The author's peer-reviewed abstract at that conference~\cite{vanrooyen2026entropy2026abstract} (``Entropy, annealing, and the continuity of agency in human--AI systems'') presented the account that became a framework called recoverable self-coding. This paper borrows some of that framework's terms and none of its claims. The study reported here is not an extended version of that abstract.}

\abstract{Brain tests of consciousness are often applied directly to large language models (LLMs). Here consciousness means access: whether information, such as a sound that was heard, becomes available for report and for use. A brain recording is in microvolts and an LLM's internal activations are arbitrary numbers, so the two share no common measurement unit. A brain test of this kind compares competing statistical models of how responses vary over repeated trials. For example, responses to a faint stimulus may vary smoothly, or they may fall into two distinct states. The test's decision names the best-fitting model. The decision is wrong when the responses truly follow a different model, for example when the test names two states for responses that vary smoothly. No method we reviewed measures how often a transferred test is wrong. We call such a method a transfer audit and propose one, the cross-entropy audit (CEA). First, our audit holds back part of the trials and fits each model to the remaining trials. It then scores how well each fitted model predicts the held-back responses, which the model has not seen. Comparing two models gives a score difference in nats per trial. The nat is a unit of information, like the bit, but based on the natural logarithm. That difference does not depend on whether the recording is in microvolts or in activations, so brain and machine are scored on one scale. Second, the audit writes down five choices that such tests make for brain data: the competing models, their fitting, what counts as one independent sample, what the uncertainty interval describes, and the decision rule. Any of these choices may need to change on a machine. Third, the audit runs the test on simulated data built to follow one model, so the right decision is known, and counts wrong decisions. In our example, the two models of a two-state test, carried over unchanged to an LLM, found two states in 989 of 12{,}000 simulated datasets that vary smoothly. The audited version of the test, adapted to the LLM, found two states in none of them. It detected 599 of 600 two-state datasets, all of them easy to detect. The audited version's uncertainty interval, however, was too narrow at six of the twelve simulated settings with smoothly varying data, and that interval is being repaired.}


\makeatletter
\def\@doiabbr{entropy}
\def\@pubvolume{0}
\def\@issuenum{0}
\def\@articlenumber{0}
\providecommand{\@datereceived}{}
\providecommand{\@dateaccepted}{}
\providecommand{\@datepublished}{}
\makeatother
\fancypagestyleassign{plainmdpi}{plain}
\fancypagestyle{plain}[plainmdpi]{\fancyfoot[R]{}}

\AtBeginDocument{\let\linenumbers\relax} 
\begin{document}
\maketitle

\section{Introduction}\label{sec:intro}
Whether an artificial system can support consciousness remains an open question~\cite{chalmers2023}. The theories discussed here use the word in two senses. The first is access. A content, meaning something perceived or thought, such as a sound that was heard, is accessed when it becomes available across a system for report and for use~\cite{mashour2020}. The second is phenomenal consciousness, meaning felt or subjective experience: whether there is anything it feels like to be that system~\cite{butlin2026}. The tests this paper concerns test predictions about access, and we call them consciousness tests only in that sense. Butlin et al.\ approach the question of consciousness in artificial systems with the \emph{indicator method}. Because no single theory of consciousness is dominant, they derive a list of indicator properties from several scientific theories and check whether a given system has each one~\cite{butlin2023,butlin2026}. Each check needs a measurement, and many of the available measurements were developed on brains. We call the move of a measurement from a brain to a system of another type a \emph{transfer}. Brain and machine measurements share no common measurement unit. Electroencephalography (EEG) records voltages on the scalp in microvolts, while the internal state of a large language model (LLM) is a vector of activations in arbitrary units fixed by training, and there is no direct conversion between the two. Browning and Veit named a broader difficulty the extrapolation problem: an indicator validated on humans needs further background assumptions before it can be used on animals or machines. For animals, structural, functional and evolutionary similarity to humans supplies those assumptions. Confidence falls as a species is more distantly related to humans. For machines, none of these similarities applies~\cite{browningVeit2020}. S\"uhr et al.\ locate the error in applying human tests to LLMs without re-establishing the measurement model and its validity~\cite{suhr2026}. Koch adds that the evidential weight of an indicator is carried over from biological cases without independent support that this weight holds in the new system~\cite{koch2026}. After a transfer, a test may therefore measure something else, and the rate at which its decision is wrong in the new system is unknown.

An unknown error rate is not a detail. A decision can enter a judgement about a system only together with the rate at which the decision is wrong. A two-state decision means little if the test also returns two states on data that are graded by construction. Here a two-state decision means that the trial-by-trial responses are best described as a mixture of two components, and graded means that they vary along a single continuum. The two states are components of a distribution of responses, not states of consciousness. The example of this paper shows how such a failure arises. In the example, the items are the concepts that the texts describe. The human test's graded model gives all items one shared threshold: the stimulus strength at which the average response rises. Where items differ in that threshold, the responses at one strength split into items already above their thresholds and items still below. A two-state mixture fits that split better than a graded model with one threshold. Carried over unchanged to an LLM, the human test's models reported two states on graded simulated data 651 times in 1{,}000 at the setting with the widest spread of thresholds (Section~\ref{sec:example-results}). Nothing in the test's output marks that decision as wrong. Simulated data from a known source can show that the decision is wrong. Their data-generating model is known, and with it whether the data are graded or two-state. Simulated data give no label for consciousness, and a machine offers no independently established labels for consciousness~\cite{koch2026}.

\begin{figure}[!t]
\begin{adjustwidth}{-\extralength}{0cm}
\centering
\includegraphics[width=\fulllength]{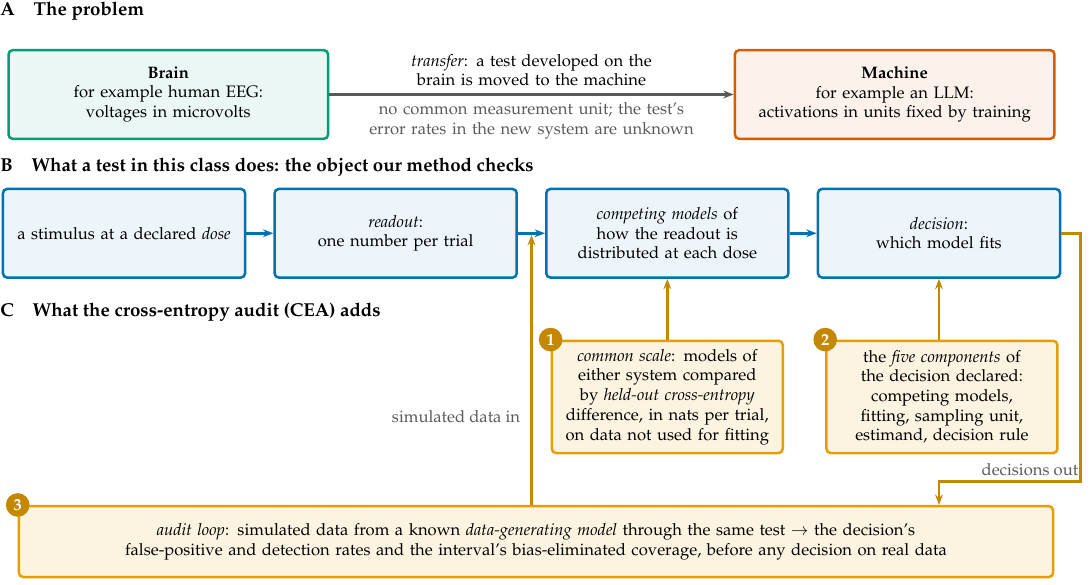}
\end{adjustwidth}
\caption{The problem (A), a test developed on the brain moved to a machine whose measurements share no unit with the brain's; what a test in this class does (B), from a stimulus at a declared dose to a decision; and what \methodname{} adds (C), a common scale for the competing models, the five components of the decision declared, and the audit loop that runs simulated data from a known data-generating model through the same test.}
\label{fig:blueprint}
\end{figure}

The aim of this paper is an instrument for a fair comparison of brain data with the measurements of an artificial or any other system. By a fair comparison we mean three things. Both systems are scored on one scale. Every choice in the test is declared. The test's error rates are measured at the new system's design before the test's decision there is trusted. The design is the new study's stimuli and its numbers of units and trials. We call any method that measures a test's error rates after its move to another kind of system a \emph{transfer audit}. We call ours the \emph{cross-entropy audit} (CEA). Both names are introduced here. The CEA applies to tests of one kind. Such a test sets a stimulus variable at random on each trial and reduces each trial to one number. It then compares competing models of how that number is distributed and returns a decision on which model fits. It has three steps (Figure~\ref{fig:blueprint}). First, it fits each competing model to most of a system's trials and asks how well the fitted model predicts the trials set aside. That prediction error, in nats per trial, is the cross-entropy. The nat is the unit of information under the natural logarithm, and one nat is about 1.44 bits. The difference between two models' cross-entropies on the same trials does not depend on whether the readout is in microvolts or in activations, so it gives one scale for brain and machine. A shared unit lets the competing models of each system be scored in the same way. The shared unit does not make the size of an effect in one system comparable with the size of an effect in the other. Second, the CEA states the five choices through which the test reaches its decision: which models compete, how they are fitted, what one sample is, what the uncertainty interval describes, and how a result becomes a decision. Each was made for the brain, and each can change meaning on a machine. Third, it runs the test on simulated data generated from one of the competing models, so that the right decision is known. It counts the test's wrong decisions before any decision on real data. We call this step the \emph{audit loop}. Section~\ref{sec:audit} gives the method in full. It defines the terms (the dose, the readout, the five components, the data-generating model) with the field's names and sources for each. It also states the hypotheses in nats and the method's scope and limits.

Other methods already set brain and artificial-network measurements side by side, and Section~\ref{sec:related} compares them with the CEA one by one (Table~\ref{tab:methods}). Representational similarity analysis compares the dissimilarity structures of the two systems' responses to shared stimuli~\cite{kriegeskorte2008}. Encoding models, among them Brain-Score, measure how well one system's responses predict the other's~\cite{schrimpf2021,brainscoreRelease2026}. Similarity indices report a number, not a decision~\cite{kornblith2019}. Behavioural benchmarks compare the errors that humans and networks make on the same trials~\cite{geirhos2020,rajalingham2018}. The psychometric literature states what a valid measurement requires and when a measurement model carries to a new population~\cite{jacobsWallach2021,suhr2026,lin2027}. Indicator frameworks assess theory-derived properties by argument~\cite{butlin2023}, or they combine behavioural, mechanistic and perturbation evidence into credences~\cite{hughesNguyen2026}. A credence is a degree of belief. Three studies of such methods, Sch\"utt et al.~\cite{schuett2023}, Avitan and Golan~\cite{avitanGolan2025} and Han et al.~\cite{han2023}, check their inference on data from a known data-generating model, within the system analysed. None of those we reviewed reports a transferred test's decision error rates in the system to which the test is moved. None reports how often the test's interval then contains an estimate of its target. The CEA measures these rates at that system's design, on simulated data, at declared settings. It leaves \emph{construct validity} to other evidence. Construct validity is whether the decision indexes the property it is meant to measure.

We demonstrate the CEA on one test, which we call the \emph{two-state access test}. The test comes from the global neuronal workspace theory. In that theory, a limited-capacity network of long-range connections, the workspace, makes a selected content available across the brain. The theory holds that a representation becomes conscious by an all-or-none event. Near the detection threshold, the responses to a stimulus of fixed strength should therefore form a mixture of two states across trials and not a single continuum~\cite{dehaene2003,mashour2020}. In the attentional blink, a second target shown shortly after a first is often missed. Behavioural ratings were bimodal at a fixed lag between the two targets, as the theory predicts~\cite{dehaene2003,sergentDehaene2004}. Sergent et al.\ tested this prediction on single-trial EEG and released their data and code~\cite{sergent2021}. We adapted the test to an LLM for a study that is held for a separate report. We call the adapted test the \emph{audited procedure} (the abstract's audited version). The theory is contested~\cite{cohen2023graded,cogitate2025}, and we take no position on it. We chose its test because the test states the distribution of single-trial responses explicitly and its data and code are public. On simulated data, the audit found very different error counts for the transferred models and for the audited procedure. On 12{,}000 datasets from twelve declared graded settings, the human test's models carried over unchanged reported two states 989 times, and the audited procedure never did. On 600 datasets from twelve declared two-state settings, the audited procedure reported two states 599 times. A replicate is one of the simulated datasets of a setting. The audited procedure's interval, however, fell short. The procedure's nominal 95\% interval contained the mean of its own statistic over the replicates (bias-eliminated coverage, Section~\ref{sec:audit-loop}) in fewer than 90\% of datasets at six of the twelve graded settings. The procedure therefore cannot yet support a confirmatory decision. These counts describe the tested settings and are not a false-positive rate in general. Section~\ref{sec:example} presents the test in full.

\textbf{Contributions.}
\begin{enumerate}
\item \textbf{The method.} We name the class of methods that measure a test's error rates after its transfer the transfer audit and propose one, the cross-entropy audit (CEA): a common scale of held-out cross-entropy differences in nats per trial, the five components of the decision declared, and the audit loop on simulated data from a known data-generating model (Section~\ref{sec:audit}, Figure~\ref{fig:blueprint}).
\item \textbf{Its place among other methods.} We compare it with methods that relate brain and artificial-system measurements, by what each puts on a common footing and whether each reports a transferred decision's error rates (Section~\ref{sec:related}, Table~\ref{tab:methods}).
\item \textbf{The example.} We apply the audit to the two-state access test (Sections~\ref{sec:example}--\ref{sec:example-results}). The work includes a re-implementation of the published human test's model comparison on the open data of twenty participants. The re-implementation reproduced the three points of the published result that this paper uses. They are the first window, at 315~ms, at which the two-state model's protected exceedance probability (Section~\ref{sec:example-human}) exceeded 0.95, the order of the models over time, and the 435--495~ms plateau. The re-implementation differed at single 30~ms windows, among them those where the effect begins. The work also includes the false-positive study on 12{,}000 graded and the detection study on 600 two-state simulated datasets (Section~\ref{sec:example-methods-calibration}). It also includes the matched comparison, a new analysis of the human data. In the matched comparison, both the graded and the two-state models let each trial's response vary in latency and in scale. On the simulated EEG sets of the brain side, the matched comparison made no two-state decision under the four graded settings. Nor did it make one under the departures O1 to O4, settings outside both fitted families. It detected 49 of 50 fitted two-state sets. Further sets were resampled from the recorded trials, so their data-generating model is unknown, and 12 of 100 of them read two-state. The graded-setting condition allowed at most 1 of 100 such sets to read two-state, so it was not met. All 50 sets of one two-state setting were unavailable, so the completion condition, which allows at most 10\% unavailable sets, was not met either. On the simulated datasets of the LLM side, the audited procedure made no two-state decision on graded data and detected 599 of 600 two-state datasets. No other procedure we ran on the same simulated data, on either side, matched both results. A repair of the interval is in development. The procedure was built for the comparison of the two systems themselves. That comparison is a further study, and no confirmatory LLM data exist. Applying the audit to a second transferred test is also further work. Nothing in this paper addresses whether either system has phenomenal consciousness, in the sense defined at the start of this section. The test audited here concerns access.
\end{enumerate}
Versions 1 and 2 of the preprint arXiv:2609.22300~\cite{vanrooyen2026accesspreprint} report an earlier version of this study, and version 3 of that preprint carries this article. Our re-implementation of the human comparison and the simulation audit (the false-positive study and detection study of Section~\ref{sec:example-methods-calibration}) reported here are the same analyses of the same saved data as in versions 1 and 2.

Section~\ref{sec:problem} states the problem of comparing brain and artificial-intelligence (AI) measurements. Section~\ref{sec:audit} presents the cross-entropy audit, and Section~\ref{sec:related} relates it to the methods of Table~\ref{tab:methods}. Section~\ref{sec:example} introduces the two-state example, Section~\ref{sec:example-methods} its methods and Section~\ref{sec:example-results} its results. Section~\ref{sec:discussion} discusses what the audit shows in general and then in the example, and Section~\ref{sec:conclusion} concludes.

\section{The Problem: Comparing Brain and AI Measurements}\label{sec:problem}

\subsection{Two Systems, Two Measurements, No Common Unit}\label{sec:problem-unit}
We call the kind of system that carries a process its \emph{substrate}: here, a brain or an LLM. The two substrates are measured in ways that share nothing. An EEG electrode records a voltage on the scalp in microvolts, the sum of currents from many neurons, and that voltage varies from one trial to the next even when the stimulus does not. An LLM processes its input in successive stages called layers. Its internal state at a layer is its \emph{residual stream}: the vector that each layer reads and adds its output to. The activations in that vector are numbers in arbitrary units fixed by training. There is no direct conversion from a microvolt to an activation.

A test developed on the brain nevertheless uses its measurement in a definite way. It presents a stimulus whose strength it has set, reduces each trial's recording to one number, and asks how that number is distributed across trials at each strength. In the test of Sergent et al., the strength is one of six sound levels of a vowel in noise, and the number is the output of a decoder trained to tell the loudest level from silence, taken in each 30~ms window~\cite{sergent2021}. An LLM offers a test of the same shape. There, the strength is the number of clues in a text, and the number is the output of a decoder at one layer. Two differences, however, lie beneath the common shape. First, a frozen LLM is a deterministic function of its input, so it has no trial-to-trial variability of its own. Whatever varies from trial to trial has to be put into the stimuli. Second, the unit that repeats is not a participant but a concept: an unnamed target about which the texts give clues. Section~\ref{sec:audit-components} calls this unit the sampling unit. The test's shape carries over. The properties of the brain that justified its choices do not.

\subsection{What a Transferred Test Carries With It}\label{sec:problem-choices}
A comparison of competing models is a procedure, and its decision depends jointly on five choices: which models compete, how they are fitted, what counts as one sample, which quantity the uncertainty interval is about, and how the result becomes a decision. In the study from which a test is carried over, each choice was justified by a property of the first substrate. In the human test audited here, those properties were a sample of twenty participants, neural variability from trial to trial in the task, and a time axis. The time axis justified the per-window decision rule. When the substrate changes, that justification lapses, even where the choice itself might still be right. Each choice therefore has to be derived again in the new application, not inherited. In practice, the transfer of the models is stated, and the other four choices travel with them without comment. The authors of the human test propose that their models may be transposed to any stimulus, task or population~\cite{sergent2021}. The cost of not assessing the other four choices can be measured, and Section~\ref{sec:example-results} reports it.

\subsection{Two Questions That Must Be Kept Apart}\label{sec:problem-two}
Two questions hide in the phrase ``does the test work on a machine''. The first is what the decision refers to: whether a two-state decision on an LLM indexes the property the test was built to detect. That is the question of construct validity, and the measurement literature supplies the tools for it. Measurement modelling makes the assumptions behind an operationalization explicit and testable~\cite{jacobsWallach2021}. The dual-validity framework asks a study for more evidence the more ambitious its claim is. It also warns of \emph{measurement phantoms}: statistical regularities mistaken for genuine phenomena~\cite{lin2027}. S\"uhr et al.\ hold that human instruments, constructed and calibrated for a human population, cannot simply be applied to LLMs~\cite{suhr2026}. Theories built on the human contrast between conscious and unconscious states may omit background conditions that an artificial system does not meet~\cite{eleos2026}. Koch separates two problems. Attributions of consciousness from indicators cannot be calibrated against independently established outcomes. And an indicator's evidential weight is carried over from biological cases without support that this weight holds across substrates~\cite{koch2026}. The calibration Koch means is the calibration of credences about consciousness.

The second question is what the procedure does: how often its decision is wrong when the data come from a known process, and how often its uncertainty interval contains the quantity it claims to. This second question has an answer that can be measured, because for simulated data the right decision is known by construction. It is the question this paper answers. We call the measurement of these two frequencies calibration, a narrower sense than Koch's. The two questions are separate, and an answer to the second does not settle the first: a rate measured against a known data-generating model cannot show what the decision refers to.

Why audit a transferred test at all, rather than build a new one for each system? Because the contrast between the two substrates is informative. A procedure's performance on human EEG does not show which of its properties survive a change of substrate. An LLM is an unusually demanding contrast: it shares no evolutionary history with the brain, no neurons, and no variability of its own from trial to trial. Suppose a procedure keeps its error rates under such a contrast, with the readout, the stimulus design and the calibration specified beforehand for the new system in its own terms. That result supports the procedure's error control at the tested settings. It is evidence about the procedure, not evidence that the procedure indexes the phenomenon.

\section{The Cross-Entropy Audit}\label{sec:audit}
This section gives the method in general terms; Figure~\ref{fig:blueprint} is its blueprint. Of the terms defined here, four are ours: the transfer audit, the cross-entropy audit, the five components as a set and the audit loop. The others are the field's, each cited where it appears: dose, readout, model, fitting, sampling unit, estimand, decision rule, held-out data, the nat, the data-generating model and bias-eliminated coverage.

\subsection{The Tests the Audit Applies To, and the Common Scale}\label{sec:audit-scale}
The CEA applies to a class of tests that share one structure. Each test sets a declared variable at random on each trial. That variable is the \emph{dose}, $K$. The test reduces each trial to a single number, the \emph{readout}. The field calls these the stimulus strength and, for a trained decoder, its projected activity~\cite{sergent2021}. The test then compares competing models of how the readout is distributed at each dose and turns the comparison into a decision. In the example of this paper, the dose is the number of clues about a concept in a text stimulus, and the readout is the output of a linear decoder at one layer of an LLM.

The audit's first step scores every competing model of either system by its cross-entropy on held-out data, in nats per trial. Held-out data are data set aside before fitting, as in cross-validation~\cite{bates2024cv}. In cross-validation the data are split into parts called folds, and each fold is held out in turn while the model is fitted to the rest. The \emph{nat} is the standard unit of Shannon information under the natural logarithm (about 1.44 bits). The readouts of a brain and an LLM still differ. What the two systems share is the scale of the score difference. Changing a readout's units shifts every model's log density by the same amount, and that amount cancels when two models are compared on the same trials. This scale compares models of a readout, not the mutual information between the two systems' representations.

The scale is built as follows. A held-out unit, a participant or a concept, contributes $n$ trials whose readouts follow a true joint law $P$, unknown to the test. A competing model $m$, once fitted to the other units, assigns those trials a predictive law $Q_m$. The model is scored by the logarithm of the probability it gave the trials it did not see, divided by $n$. The expectation of that score under $P$ is minus the joint \emph{cross-entropy} of the unit's $n$ trials divided by $n$, where $H(P, Q_m) = -\,\mathbb{E}_P[\log Q_m]$ and $\mathbb{E}_P$ is the expectation over trials drawn from $P$. The test's statistic is the difference between the held-out scores of two models, 1 and 2, with predictive laws $Q_1$ and $Q_2$, and its expectation, conditional on the fitted models, is a difference of cross-entropies per trial,
\begin{equation}
\Delta = \frac{H(P, Q_1) - H(P, Q_2)}{n} = \frac{D_{\mathrm{KL}}(P \,\|\, Q_1) - D_{\mathrm{KL}}(P \,\|\, Q_2)}{n},
\label{eq:ce}
\end{equation}
where $D_{\mathrm{KL}}(P \,\|\, Q)$ is the Kullback--Leibler divergence of $Q$ from $P$, the expected excess log loss of predicting with $Q$ in place of $P$. The second form follows because a cross-entropy is the entropy of $P$ plus that divergence, and the entropy of $P$ cancels in the difference. Equation~\eqref{eq:ce} is in nats per trial, and a positive value favours model~2. The log score is a proper scoring rule: on average it is highest for the true law. Because the scored trials were not used in fitting, a more flexible model scores better only when it is closer to the true law, not because it fits the training trials more closely. That a scoring rule is proper is a statement about expectations. It gives no guarantee about selection among a finite set of fitted models on one dataset. How that selection behaves on one dataset is what the audit loop of Section~\ref{sec:audit-loop} measures. Where a model includes a random effect for the unit, the score is the joint predictive probability of the unit's trials with that effect integrated out, and not a sum of single-trial terms.

A common scale is not a common effect. The two applications of a transferred test differ in readout, design and competing models, so they define different predictive tasks, and a difference of a given size in equation~\eqref{eq:ce} need not mean the same thing in both. The unit makes the two results reportable beside one another. It does not license reading one against the other as a magnitude, and the audit never compares the brain's value of $\Delta$ with the machine's.

\subsection{The Five Components of a Decision}\label{sec:audit-components}
The audit's second step declares the five \emph{components} of the test's decision. We call the five choices of Section~\ref{sec:problem-choices} by that name. They are the parts of the statistical decision problem that lead from the data to a decision. Each part keeps its standard name. The \emph{competing models} are the candidate statistical models of how the readout is distributed at each dose. They are grouped into families of models of one form (in the example, graded and two-state). Their estimation, including any model selection, is the \emph{fitting}~\cite{wilsonCollins2019}. The \emph{sampling unit} is the unit over which the result generalizes, such as the participant in a human study. The \emph{estimand} is the quantity about which the uncertainty interval makes a claim~\cite{morris2019}. The \emph{decision rule} converts the statistic and its interval into a decision. Each of the five was justified by a property of the first system, and each can change meaning in the second.

Why these five and not others? Each corresponds to one part of a statistical decision problem. The competing models are the hypothesis space, the models of the readout that are allowed to compete. The fitting is the estimation map, from data to a fitted member of each family, with the optimizer and the model selection nested inside it. The sampling unit is the sampling model: which units are treated as exchangeable, that is, as interchangeable repetitions, and therefore which kind of repetition the uncertainty refers to. The estimand is the quantity, defined under that sampling model, that the interval makes a claim about. The decision rule is the decision function, from the statistic and its uncertainty to the reported decision. Once these five, with the readout, the stimulus design and the numerical implementation, are fixed, the decision is determined by the data. The list describes one statistical decision problem in the estimand vocabulary of simulation-study planning~\cite{morris2019}, and it is the set the audit measures. Three declarations precede the five and are not counted among them: the readout, the stimulus design and the numerical implementation. They fix the design on which the five statistical choices are then assessed. Fold construction and model selection enter through the fitting. Dependence between repeated measurements enters through the sampling unit, and the randomness of a fitting algorithm is either fixed by its seeds or included in that sampling model. A seed is the starting value of a random-number generator. A procedure that reports no interval declares which uncertainty claim is absent.

The sampling unit deserves a word on its own, because the transfer changes it. The choice is forced by where the randomness comes from: an interval generalizes over whatever is treated as exchangeable. In a human study the unit is the participant, and the inference concerns new participants, because each participant is a new draw from a population of brains with variability of its own from trial to trial. A \emph{checkpoint} is an LLM with fixed weights, its trained parameters. A fixed checkpoint is one system and a deterministic function of its input. There is no population of systems to sample at one checkpoint, and no trial-to-trial noise in the substrate. Retrained instances could supply a population, as in comparisons of trained vision networks~\cite{rajalingham2018}, but not for a released pretrained checkpoint. The variability therefore has to be placed in the stimulus by design, and the unit becomes whatever the stimulus design repeats. In the example it is the concept, and the inference concerns new concepts of the same kind. That population is small: the concepts come from published lists of the properties people associate with concepts, and the study uses 91 of the 94 that qualify (\SMdesign). Whether an LLM is one participant or many, and which unit carries the inference, has to be decided on principle~\cite{suhr2026,binz2023}. Treating the stimulus unit as the sampling unit is an assumption the transfer introduces and the human study never had to make, and the audit loop checks it against replication (Section~\ref{sec:audit-loop}).

\subsection{The Audit Loop}\label{sec:audit-loop}
The audit's third step passes simulated data through the same procedure. Their data-generating model~\cite{morris2019}, the process that produced them, is known by construction, and with it the family that the model belongs to (in the example, graded or two-state). The audit therefore measures how often the decision names the wrong family, how often it detects the right one, and how often the interval contains the mean of the estimates of its target, all before any real-data decision. This pass is the audit loop of Section~\ref{sec:intro}. The loop applies the logic of two checks from the statistical literature to a test after its transfer. The first is model recovery, in which data simulated from each candidate model are fitted by every candidate, to see whether the generating model is recovered~\cite{wilsonCollins2019}. The second is the simulation study, which evaluates a statistical method on data generated under a known mechanism~\cite{morris2019}. The loop is the check that remains when the new system has no independently established labels for consciousness~\cite{koch2026}: an error rate can be measured only on data whose right decision is known. A moved test can change meaning at any of the five components while its output looks no different.

The data are simulated because the data-generating model has to be known, so that every decision can be scored against it, and no real recording supplies that, on either substrate. The simulated readouts take the place of the new system's readout, at the new study's design sizes. The design sizes are the new study's numbers of sampling units, trials, layers and folds. The rates the loop measures are therefore properties of the procedure at that design. The loop does not simulate the system itself, its decoders or its stimuli. A \emph{generator} $g$ is a data-generating model of this kind: one member of a competing family at declared parameter values, used to simulate data at the test's declared design sizes. The generators are chosen from the family that the procedure is most likely to mistake for its hypothesis. Their settings carry declared mechanisms by which that mistake can arise. In the example these are three mechanisms in the data-generating model, chosen for this study, by which a one-state process can produce apparently all-or-none structure. That a single continuum can yield distributions read as all-or-none is shown in~\cite{cohen2023graded}. Each mechanism has a zero setting beside the base generator, the member inherited from the human comparison (Section~\ref{sec:example-families}), so that the effect of the mechanism is separated from the effect of changing the member. Such data are adversarial within the hypothesis space: they come from the family that the hypothesis competes against and are chosen to be hard to tell from the hypothesis. They are not adversarial to the fitted family itself.

Two rates of the decision are kept apart, because each is defined over a different set of data. Let $R$ replicate datasets be drawn under a generator $g$, and let the procedure return on replicate $r$ a decision $d_r$, an interval $[\ell_r, u_r]$ and a point estimate $\bar\Delta_r$ of its statistic. Where $g$ belongs to the family the test's hypothesis competes against, a decision for the hypothesis is false with respect to the data-generating model, and the \emph{false-positive rate} is
\begin{equation}
\hat F_g = \frac{1}{R} \sum_{r = 1}^{R} \mathbf{1}\{d_r \text{ names the hypothesis}\},
\label{eq:fpr}
\end{equation}
where $\mathbf{1}\{\cdot\}$ is one when its argument holds and zero otherwise, every attempted replicate enters the denominator, and a replicate that returns no decision counts as no false positive. Where $g$ belongs to the hypothesis's own family, the same count is the \emph{detection rate}. The two are never pooled: a false-positive rate is defined only over data from the competing family, and a detection rate only over data from the hypothesis's family. A false positive under a graded generator is false with respect to that data-generating model; it is not a type-I error against a null of zero expected score difference, because a misspecified model's predictive preference can be positive in truth. The decision says which family predicts held-out data better. The audit counts it as false when that family did not generate the data, because the decision is used to name the generating family. When no false positive is observed in $R$ replicates, the exact one-sided 95\% binomial upper bound on the false-positive probability is
\begin{equation}
F^{+} = 1 - 0.05^{1/R},
\label{eq:binom}
\end{equation}
which is 0.003 at $R = 1{,}000$. A zero count in equation~\eqref{eq:fpr} supports the bound of equation~\eqref{eq:binom} and nothing stronger.

The interval is checked separately, against its own target. The target of the interval is the estimand, $\theta_g$. It is the expectation of the complete procedure's statistic under $g$ at the design sizes, taken over the stimulus draws, the folds and the optimizer's starts. The estimand is a cross-validated quantity and therefore an average over possible training sets, which is what cross-validation estimates~\cite{bates2024cv}. It is neither the expected loss of one fitted model nor a property of the generator alone. Because $\theta_g$ is not known in closed form, the audit estimates it from the replicates themselves and counts how often each interval contains the estimate:
\begin{equation}
\hat\theta_g = \frac{1}{R_v} \sum_{r \in \mathcal{V}} \bar\Delta_r, \qquad
\hat C_g = \frac{1}{R_u} \sum_{r \in \mathcal{U}} \mathbf{1}\{\ell_r \le \hat\theta_g \le u_r\},
\label{eq:cov}
\end{equation}
where $\mathcal{V}$ is the set of the $R_v$ replicates whose fits and point estimate are numerically valid and $\mathcal{U} \subseteq \mathcal{V}$ is the set of the $R_u$ replicates with a usable, finite interval. $\hat C_g$ in equation~\eqref{eq:cov} is the \emph{bias-eliminated coverage} of Morris et al.~\cite{morris2019}: the share of usable intervals that contain the mean of the estimates. $\hat C_g$ is not a coverage probability, which is the probability that an interval contains its true target, because the reference is estimated from the same replicates, and the plug-in binomial standard error $\{\hat C_g (1 - \hat C_g) / R_u\}^{1/2}$ treats $\hat\theta_g$ as fixed and so does not represent that dependence. Where the estimate of $\theta_g$ is itself unstable at a setting, the reference of the bias-eliminated coverage there is unstable as well. Bias-eliminated coverage is nevertheless the check an audit of error rates alone omits: a procedure can make no false positive and still carry an interval that misses its own target, and the shortfall appears only when the estimand is named and each interval is checked against the mean of the estimates.

Every frequency measured in the loop, equations~\eqref{eq:fpr} and~\eqref{eq:cov}, holds at the settings and design sizes at which it was measured. The generators, their settings and the number of replicates are the audit's own design, so the totals describe behaviour at those settings and are not an error rate of the procedure in general. One thousand replicates per setting put the Monte Carlo standard error near 0.007 at a reference rate of 0.05. The Monte Carlo standard error is the uncertainty that comes from running a finite number of simulated datasets. The audit states a calibration rule for each frequency before the loop runs. In the example the calibration rules are a false-positive rate of at most two standard errors above 0.05 and a bias-eliminated coverage of at least 0.90 at every setting. The loop also checks the sampling unit. The spread of the statistic across replicate datasets at one setting is the uncertainty that the unit should claim. The ratio of that spread to the claimed standard error says whether the standard error claimed under that unit matches the replicate spread there.

\subsection{Hypotheses Stated in Nats}\label{sec:audit-nats}
Hypotheses can be stated in the same units as the score. With $K$ the dose, let $S$ be the latent state that a model assigns to a trial. The \emph{mutual information} $I(K;S)$, in nats, is the information that the state carries about the dose, and the \emph{entropy} $H(S)$ is the information needed to specify the state. A model that assigns each trial to one of $N$ latent states then satisfies $I(K;S) \le H(S) \le \ln N$, so two states carry at most $\ln 2$ nats, one bit, about the dose. This bound is set by the number of states and is not a measured capacity. These are Shannon quantities of readout distributions, and we estimate no thermodynamic entropy for either system.

The bound places the hypothesis inside a decomposition of the information that the readout carries. Let $Y$ be the readout. Under a model with a latent state, the dose reaches the readout through two routes: the state, and the readout's distribution within each state. The chain rule of mutual information, $I(K; S, Y) = I(K; S) + I(K; Y \mid S) = I(K; Y) + I(K; S \mid Y)$, gives
\begin{equation}
I(K; Y) = I(K; S) + I(K; Y \mid S) - I(K; S \mid Y) \;\le\; I(K; S) + I(K; Y \mid S),
\label{eq:decomp}
\end{equation}
where $I(K; Y \mid S)$ is the information about the dose carried by the readout within a state and $I(K; S \mid Y)$ is the information about the dose that the state holds beyond the readout. The last term vanishes when the dose and the state are conditionally independent given the readout. In the two-state hypothesis of the example, the first term is at most $\ln 2$ nats and the second term is the information in the location of the responsive, high state (Section~\ref{sec:example-families}). In the graded alternative there is one continuous state and no such split, so all of $I(K; Y)$ is carried by it. The two families of a test are statistical operationalizations of these two claims, with additional distributional assumptions, and predictive preference between them does not prove that the underlying state is binary or continuous.

Three negative statements go with the decomposition. The decomposition is a consequence of the latent-state representation and not a test of a latent state. The competing-model comparison is the test. The decomposition's terms describe the channel from dose to readout that a fitted model defines, and they are distinct from the predictive comparison of equation~\eqref{eq:ce}: equation~\eqref{eq:decomp} does not decompose equation~\eqref{eq:ce}. And every quantity in it is a Shannon quantity of a conditional readout distribution. Both substrates are organized ordinary matter held far from equilibrium by a throughput of energy, and that shared condition motivates this paper's information-theoretic framing. The audit, however, supplies no thermodynamic reference state and estimates no thermodynamic entropy. The audit's quantities are extrinsic Shannon quantities of the channel from the dose to the readout, in the sense that Tononi et al.\ distinguish from the intrinsic, causal information of integrated information theory~\cite{tononi2016}, and none of them estimates integrated information. A two-state model is a statistical coarse-graining of the readout's law. It is not evidence that the system performs a many-to-one operation on its information-bearing state, because information absent from one projection can remain elsewhere~\cite{sagawa2014}, and any erasure accounting depends on what is eventually reset and what side information remains~\cite{wolpert2019}. Equation~\eqref{eq:ce} compares two predictive distributions for the same observations under a third distribution, the true law $P$, and is not an entropy-production identity, which compares physical forward and reverse path measures under a specified protocol~\cite{seifert2012}. We therefore estimate neither heat dissipation nor entropy production for either substrate, and no inference is made from any of this to physical erasure, heat or entropy production. \SMentropy{} gives the detail behind this boundary.

\subsection{Scope and Limits}\label{sec:audit-scope}
The method is general in design, and we demonstrate it on one test. We do not show that it carries to another test, and we do not claim that the five components suffice for every test. A test settled by inspecting a system's architecture, with no probabilistic readout, is outside the class, and applying the declarations to it would require a quantitative competing-model test with its own design and calibration. Two further indicators fall inside the class. One is \emph{metacognitive sensitivity}, called meta-$d'$ and estimated from choices and confidence ratings under a signal-detection model~\cite{maniscalco2012,servajean2026}. The other is the aggregate ignition index, where ignition is the theory's name for the late, all-or-none entry of a content into the workspace (Section~\ref{sec:example-prediction}); the index fits candidate curves along a named axis to the accuracy of a trained probe (a classifier trained on internal activations)~\cite{rahbar2026}. For both we declare the components but do not run the audit loop: \SMdeclarations{} sets out, for each, the competing models, fitting, sampling unit, estimand, decision rule, the declarations made upstream of the five, and the checks an audit would require. Those declarations specify requirements for further audits. We did not audit either indicator's estimation pipeline or interval behaviour, and the declarations do not claim that either published measure has been shown false.

The audit's output is also limited in what it can settle. A false-positive rate measured against a known data-generating model shows how often the procedure names the wrong family at the tested settings. Recovering the family of a data-generating model is not the same as showing that the decision indexes the property the test was built to detect. That is the construct-validity question of Section~\ref{sec:problem-two}. A neural index needs validation through the property itself~\cite{kouider2010}, and no rate measured in simulation answers that question. The decisions audited here concern which model of a readout fits, a functional question; whether either system has phenomenal consciousness is a further question, and no result here addresses it.

\section{Relation to Other Methods}\label{sec:related}
Other methods put brain and artificial-network measurements on a common footing, and Table~\ref{tab:methods} compares them with the CEA by three questions: what each puts on a common footing, what it measures, and whether it reports the error rates of a transferred decision in the system to which the test was moved. Section~\ref{sec:related-comparison} takes the methods one by one. Section~\ref{sec:related-parents} gives the statistical parents of the audit loop, and Section~\ref{sec:related-construct} the measurement literature that owns the construct question. Section~\ref{sec:related-adds} states what the audit adds and what it leaves to other evidence.

\subsection{Methods That Compare Brain and Artificial-Network Measurements}\label{sec:related-comparison}
Representational similarity analysis summarizes each system by the matrix of dissimilarities between its responses to every pair of a shared set of stimuli, and it compares the two matrices, so that no map between neurons and units is needed~\cite{kriegeskorte2008}. Sch\"utt et al.\ gave its model comparisons an inference that generalizes to new participants and new stimuli, and they validated that inference on simulated and resampled data from a known data-generating model, within the system analysed~\cite{schuett2023}. Their simulations also measured the cost of a mis-declared sampling unit. Tests that treated the participant as the only random factor rejected a true null hypothesis in up to 60\% of simulations when the stimuli had in fact been sampled as well. Encoding models fit a map from a network's activations to neural responses and score how well the fitted map predicts held-out responses, often relative to the reliability of the neural data~\cite{schrimpf2021,caucheteux2022}. Brain-Score, a public platform, scores hundreds of artificial networks against dozens of neural and behavioural benchmarks~\cite{brainscoreRelease2026}. Such comparisons measure correspondence, the agreement between the two systems' responses, for the tasks and scores declared. Testing more systems does not establish the error rates of a transferred decision. Centered kernel alignment scores how alike two representations' similarity structures over shared examples are, without fitting a map, and it reports a similarity between 0 and 1, not a decision~\cite{kornblith2019}. Behavioural benchmarks compare the errors that humans and networks make on the same trials~\cite{geirhos2020,rajalingham2018}. For one of them, Geirhos et al.\ found that the approximate interval in common use for their agreement statistic was wrong, and they replaced it with bands simulated under a declared null of independent observers~\cite{geirhos2020}.

Two recent studies turned the recovery logic on the comparison metrics themselves. Avitan and Golan ran model recovery (Section~\ref{sec:audit-loop}) for a comparison of networks with human behaviour: each candidate network in turn generated simulated judgements, every candidate was refitted to them, and the study recorded how often the generating network won. With a fitted linear map, the network that had generated the simulated judgements was recovered in fewer than 80\% of simulations, even with 4.2 million training triplets of images~\cite{avitanGolan2025}. Han et al.\ replaced the brain with a known artificial network as the target and asked whether the linear encoding model and centered kernel alignment identify its architecture. For several target layers the best-matched candidate had the wrong architecture, and identification depended on the stimulus set and on the number of units recorded~\cite{han2023}. Both studies make the right answer known by construction, and so does the audit loop. Han et al.\ make it known by changing the system: they put a known network in the brain's place. Avitan and Golan make it known for a selection metric, in the domain where the metric is used. Each study measures a metric's error within one domain, and neither moves a test to a new system. The audit loop makes the right answer known through the data. Simulated readouts from a known data-generating model pass through the test on the system to which the test was moved. What is counted is that test's decision error after its transfer.

\begin{table}[!t]
\begin{adjustwidth}{-\extralength}{0cm}
\caption{Methods that relate brain and artificial-system measurements, by what each puts on a common footing, what it measures and whether it reports a transferred decision's error rates in the new system, with within-system validation noted where the sources report it.\label{tab:methods}}
\footnotesize
\begin{tabularx}{\fulllength}{>{\raggedright\arraybackslash}p{0.19\fulllength} L L >{\raggedright\arraybackslash}p{0.25\fulllength}}
\toprule
\textbf{Method} & \textbf{What it puts on a common footing} & \textbf{What it measures} & \textbf{Error rates of a transferred decision in the new system?} \\
\midrule
Representational similarity analysis~\cite{kriegeskorte2008,schuett2023} & The geometry of dissimilarities between responses to a shared set of stimuli & How well each candidate network's dissimilarity structure predicts the brain's & No transfer; inference validated within the analysed system on simulated and resampled data~\cite{schuett2023} \\
\addlinespace
Encoding models and Brain-Score~\cite{schrimpf2021,caucheteux2022,brainscoreRelease2026} & Shared stimuli, through a fitted map from network activations to neural responses & Held-out predictive correlation, often relative to the reliability of the neural data & No \\
\addlinespace
Centered kernel alignment~\cite{kornblith2019} & Similarity structures over shared examples & A similarity between 0 and 1 & No decision \\
\addlinespace
Behavioural benchmarks (error consistency)~\cite{geirhos2020,rajalingham2018} & The same trials or images, with correctness coded alike & Agreement of errors, or of item-level difficulty & No; the agreement statistic's interval checked against null bands simulated from independent observers~\cite{geirhos2020} \\
\addlinespace
Model recovery of comparison metrics~\cite{avitanGolan2025,han2023} & Candidate networks against the same data: human behaviour, or a known network standing in for the brain & Whether the generating network is selected, or the target network's architecture identified & A metric's error within one domain; no transfer \\
\addlinespace
Measurement validity and invariance~\cite{jacobsWallach2021,suhr2026,lin2027} & A measurement model required to hold in both populations & Whether the measurement model holds; what a claim must show & No; requirements, not a measured rate \\
\addlinespace
Indicator frameworks~\cite{butlin2023,butlin2026,hughesNguyen2026} & Theory-derived indicator properties & Whether a system has each property; credences & No; error rates of a transferred decision not reported; properties assessed by argument or by scored evidence \\
\addlinespace
Simulation-based calibration~\cite{talts2018} & None: one assumed model & Whether an algorithm's posterior samples are correct & No; posterior computation, not decisions \\
\addlinespace
The cross-entropy audit, CEA (this paper) & The scale on which competing models of each system's readout are scored: held-out cross-entropy differences in nats per trial & False-positive and detection rates of the decision; how often its interval contains the mean of the estimates of its target & Yes, at the new system's design, on simulated data from a known data-generating model, at declared settings \\
\bottomrule
\end{tabularx}
\end{adjustwidth}
\end{table}

\subsection{The Statistical Parents of the Audit Loop}\label{sec:related-parents}
The audit loop is not a new idea in statistics, and Section~\ref{sec:audit-loop} named its two parents. The first is model recovery (Section~\ref{sec:audit-loop}). Its full result is a \emph{confusion matrix}, the frequency with which data from each generating model are reported as each candidate~\cite{wilsonCollins2019}. Wilson and Collins place the check before any fit to real data, and they warn that its result depends on the parameter ranges simulated and that a model comparison ranks only the models considered. The audit loop applies this logic at the level of model families, at declared settings. In a two-by-two table of generating family against decided family, the false-positive rate is the graded-to-two-state cell and the detection rate the two-state-to-two-state cell, each measured only at the generators declared.

The second parent is the simulation study (Section~\ref{sec:audit-loop}), whose aims, data-generating mechanisms, estimands, methods and performance measures are stated before it runs. Morris et al.\ define the performance measures the audit uses, among them coverage and the rejection rate, give each its Monte Carlo standard error, and recommend that the number of repetitions be chosen from that error~\cite{morris2019}. The false-positive study's thousand replicates per graded setting were chosen that way (Section~\ref{sec:audit-loop}). The audit's estimand is the expectation of its own statistic, and its reference is the mean of the estimates. That is why equation~\eqref{eq:cov} is bias-eliminated coverage, which Morris et al.\ list beside coverage of the true value as a diagnostic, not as a performance measure in its own right~\cite{morris2019}. Because the audit's target is the expectation of its own statistic, the procedure has no bias with respect to it, and bias-eliminated coverage is here the coverage of that target up to the error in its estimate. Within brain-model comparison, Sch\"utt et al.\ are the nearest precedent for the loop. They simulated data from a known generating model, ran the complete inference and measured its type-I error rates and the accuracy of the uncertainty estimate~\cite{schuett2023}.

Simulation-based calibration shares a word with the audit's calibration and little else. It draws many parameter values from a prior, the distribution assumed for the parameters before the data, and simulates a dataset from each. It then checks that the rank of each true value among the draws from its posterior, the distribution of the parameters given the data, is uniform~\cite{talts2018}. A uniform rank distribution validates the computation of a posterior under one assumed model, and the method assesses no model-comparison decision. One of its lessons carries over. A single simulated truth is a blunt instrument, and a method is checked over the range of truths that matter. The audit loop follows that lesson by running over a grid of declared settings and reporting each setting's rates on their own, not an average over a prior.

Each parent validates a method in the setting where the method is used. The audit runs the same logic on a test after its move to a system of another type, with the five components declared first. If the rates change after the transfer, the five components name the assumptions to investigate. The contribution of each component, however, may not be separable from the others. Two caveats come with the parents and apply here. The candidate set is closed: recovery assumes that the true generator is among the candidates, and Avitan and Golan call closed-set recovery necessary but not sufficient~\cite{avitanGolan2025}. The true generator of an LLM's readout need not belong to either declared family, so a decision for one family is a decision relative to the declared alternatives. And a recovery rate depends on the stimulus set and the amount of data~\cite{avitanGolan2025,han2023}, which is why every frequency of Section~\ref{sec:audit-loop} is reported at its tested settings and design sizes only.

\subsection{The Construct Side: Measurement Validity and Indicator Frameworks}\label{sec:related-construct}
Section~\ref{sec:problem-two} separated what a decision refers to from what the procedure does. The measurement literature owns the first question, and it states requirements, not measured rates. S\"uhr et al.\ set out how a psychometric instrument is built: a construct is modelled, items are written, the instrument is calibrated on a target population, and its scores are interpreted within that population~\cite{suhr2026}. They name two ways a transfer to LLMs can fail. The construct may not exist or may differ in the new population, and the item parameters calibrated on humans need not carry over. The condition they require for any comparison between populations is \emph{measurement invariance}: the distribution of the responses given the latent trait, the unobserved quantity the instrument is meant to measure, does not depend on which population the respondent comes from~\cite{suhr2026}. Their illustration is a personality item and its loading, the strength of the item's relation to the trait. On an LLM the loading was less than half of its value in a human sample. In their view, the error is a transfer that does not re-establish the measurement model and its validity. For one competing-model test, declaring the five components re-establishes the procedural side of the measurement model. The second of their two failures, parameters calibrated on one population and inherited by another, is the kind of failure the audit measures.

Measurement modelling brings construct validity from the quantitative social sciences in several aspects, among them whether a measure agrees with other measures of the same construct, differs from measures of other constructs, and predicts what the construct should predict~\cite{jacobsWallach2021}. In its terms the audit is evidence about the decision procedure and bears on none of these aspects. The dual-validity framework joins that tradition with the standards of causal inference, and it asks a study for more evidence the more ambitious its claim is. It also names threats to the conclusions drawn from an LLM's outputs, among them treating repeated completions as independent participants and borrowing statistical defaults from human research~\cite{lin2027}. In our reading, those two threats are qualitative statements of two components, the sampling unit and the decision rule. Both frameworks describe what a good measurement requires. Neither states what a particular procedure's error rate becomes when it is carried to a new substrate, because answering that needs data whose data-generating model is known.

Indicator frameworks sit on the same side. An indicator claim has two parts: that the system has the property, and that the property raises the probability that the system is conscious, relative to a theory~\cite{chalmers2023,butlin2026}. Butlin et al.\ assess the first part by inspection of a system's architecture and training against theory-derived definitions, and they treat each indicator as shifting credence, a degree of belief, without assigning the indicator a probability~\cite{butlin2023,butlin2026}. Hughes and Nguyen's triangulated assessment stack combines behavioural, mechanistic, perturbation and observer-control evidence into credence reports, with its thresholds set by justification~\cite{hughesNguyen2026}. The audit bears on the measurement in the first part of such a claim, where the property is assessed by a quantitative test with a decision, and on nothing in the second. Where such a framework takes its evidence from a test of that kind, an audit of the test supplies what the framework otherwise lacks: the decision's measured error rates at the tested settings. A newly developed AI-specific test of the same quantitative competing-model kind has the same five components and requires the same evidence. For such tests, the five components are a precondition for building AI-specific instruments accountably, and not a rival to building them.

\subsection{What the Audit Adds, and What It Leaves to Other Evidence}\label{sec:related-adds}
Each of the eight reviewed methods of Table~\ref{tab:methods} measures correspondence, validates inference or states requirements, with no decision moved to and measured in a second system. None reports a transferred test's decision error rates in the system to which the test is moved. That is the gap named in Section~\ref{sec:intro}. The CEA measures those rates on simulated data at declared settings. It adds two things to its statistical parents. It measures the error rates after the transfer, at the design of the system to which the test was moved. And it declares the five components beforehand. What it leaves to other evidence is construct validity, whether the decision indexes the property the test was built to detect (Section~\ref{sec:problem-two}), and with it the second part of an indicator claim. The comparison is a statement about the works reviewed, not about what any method could be made to do. The tools behind several rows of Table~\ref{tab:methods} accept data from any system, and the statement concerns what the methods report and require.

\section{The Example: Two-State Access}\label{sec:example}
\begin{figure}[!t]
\begin{adjustwidth}{-\extralength}{0cm}
\centering
\includegraphics[width=\fulllength]{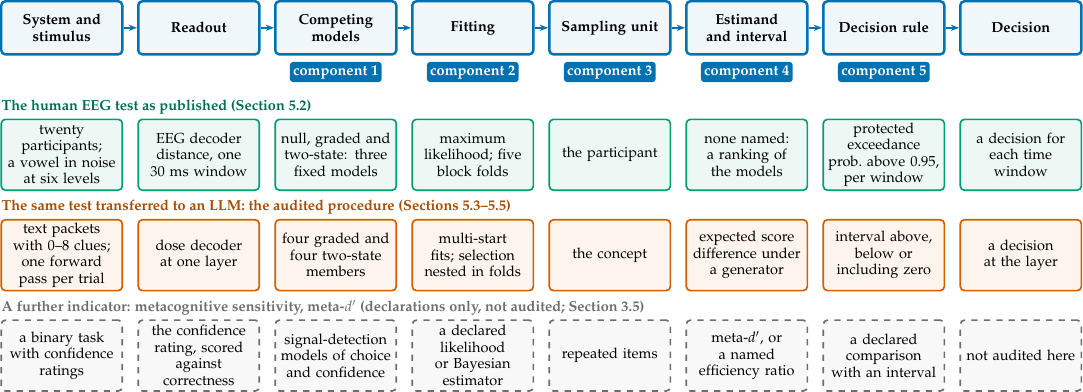}
\end{adjustwidth}
\caption{The stages of a test in the audited class (Figure~\ref{fig:blueprint}B) and the five components, instantiated by the human EEG test as published (top), by its transfer to an LLM, the audited procedure (middle), and by metacognitive sensitivity, meta-$d'$, declared but not audited (bottom, dashed).}
\label{fig:instances}
\end{figure}

The method is general, and this section introduces the one test on which we demonstrate it: the prediction it tests and the accounts that compete over it (Section~\ref{sec:example-prediction}), the human test as published (Section~\ref{sec:example-human}), its adaptation to an LLM, the audited procedure (Section~\ref{sec:example-adaptation}), the two families of competing models (Section~\ref{sec:example-families}) and the statistic and the decision rule (Section~\ref{sec:example-statistic}). Figure~\ref{fig:instances} sets the published test and its transfer side by side, stage by stage. The methods and results of the example follow in Sections~\ref{sec:example-methods} and~\ref{sec:example-results}.

\subsection{The Prediction and the Competing Accounts}\label{sec:example-prediction}
The global neuronal workspace theory, proposed as an architecture of many specialized processors and one capacity-limited workspace built from long-range neurons~\cite{dehaene1998pnas,dehaeneNaccache2001}, concerns \emph{access}: the entry of a representation into that workspace, from which it is broadcast to the rest of the system and becomes reportable~\cite{mashour2020}. Its dynamical claim is that entry is a late, all-or-none event, called \emph{ignition}. Near the detection threshold, the responses to a stimulus of fixed strength should then form a mixture of two states across trials (two-state) and not a single continuum (graded)~\cite{dehaene2003,sergentDehaene2004}. The prediction has a history. A spiking model of the attentional blink, the failure to report a second target shown shortly after a first, produced firing rates in higher areas that were ``distributed bimodally across trials''~\cite{dehaene2003}. The model's authors warned that gradual reportability ``may be an artificial consequence of averaging across trials''. In the companion behavioural experiments, visibility ratings at a fixed lag between two targets in rapid succession were bimodal, while ratings under masking alone were unimodal and shifted continuously with target duration~\cite{sergentDehaene2004}. Averaged evoked responses show a late nonlinear threshold for access~\cite{delCul2007}, and a later review states that the theory's simulations lead to ``a bimodal `all-or-none' distribution of activity''~\cite{dehaeneChangeux2011}. The claim concerns entry, and an accessed content can still vary in intensity~\cite{dn2026}. It concerns access to one content, and not transitions in an organism's overall state of consciousness, such as anaesthesia, which are a different observable~\cite{mashour2020,dehaeneChangeux2011}.

An average response cannot decide the claim, because at one stimulus strength a unimodal response and a mixture of weak and strong responses can have the same mean~\cite{sergent2021,windey2014}. A nonlinear mean is compatible with graded awareness~\cite{timmermans2010}. Awareness is the cited authors' word for conscious perception as participants report it. Averages conditioned on a binary report, such as seen or unseen, can show an ignition-like step that arises from a graded inference~\cite{fleming2020}. The test therefore compares the distributions of single-trial responses against graded alternatives that allow a nonlinear mean and a changing variability, and it does not sort trials by report. Two fitted components need not produce two visible peaks, and a readout with two peaks does not by itself identify recurrent circuitry.

Evidence for discreteness began with report and then moved to neural readouts, and critics answered with evidence of the same kinds. \SMaccounts{} records the positions of ten accounts in their authors' terms, and how a two-state or a graded result on this readout would bear on each, without adjudicating between theories. Only the workspace account states a prediction of a two-state distribution across trials for this readout~\cite{dehaene2003,dehaeneChangeux2011,sergent2021,dn2026}. Two reconciliations predict that the answer depends on the level of representation. Partial awareness holds that access is all-or-none within each level and graded only in how many levels of a content are accessed~\cite{kouider2010}. Levels of processing holds that awareness is graded for low-level features and all-or-none for high-level ones~\cite{windey2014,windeyCleeremans2015}. Continuum signal detection places report on a single continuum of representational strength and allows that entry into awareness may itself be all-or-none~\cite{cohen2023graded,sergent2021}. A two-state result would therefore be in tension with continuum signal detection only if the readout indexes the strength of the content. In predictive processing the hypothesis with the highest posterior probability is selected for conscious perception without a separate threshold mechanism~\cite{hohwy2012}. The predictive workspace keeps an ignition threshold within predictive processing~\cite{whyteSmith2021}. Each of these readings assumes that the readout indexes the account's object at the stage the account concerns. A flexible graded model that loses on this readout would not refute continuum theories in general, and graded support would not establish an absence of recurrence. The theory is contested on other grounds as well. Behavioural studies of graded awareness dispute the two-state picture~\cite{cohen2023graded}. In an adversarial test, the proponents of rival theories agree on opposing predictions before the data are collected. The largest adversarial test to date, on 256 participants, challenged two of the theory's preregistered predictions about suprathreshold stimuli, those presented above the detection threshold~\cite{cogitate2025}. That test registered no prediction about the near-threshold distribution, which is the signature examined here~\cite{melloni2023}. We use the theory's test for its explicit prediction about single-trial responses and its public data and code, not for the theory's standing, and we take no position on the theory.

\subsection{The Human Test}\label{sec:example-human}
Why this test, and why this dataset? The audit needs a published test whose components it can take over exactly, so that the transfer it audits is the one that would be made: the test's models, fitting, sampling unit and decision rule (the published test named no estimand), and the data and code with which its decision can be reproduced. Sergent et al.\ released all of these, and they propose that their models may be transposed to any stimulus, task or population, so the paper audits the transfer they invite~\cite{sergent2021}. Their test also states the prediction as a distribution of single-trial neural responses, which is the signature the audit needs, and the largest adversarial test of the theory registered no prediction about that distribution (Section~\ref{sec:example-prediction}). The twenty participants are the published study's sample and not a benchmark of ours. The audit's error rates come from simulated data from a known data-generating model and do not depend on that sample size. The human data serve two purposes: to reproduce the published decision, which Section~\ref{sec:example-results} reports as reproduced on three points, with its sensitivities, and to supply the components that were carried over. The human result is a reference, not a standard. A second open dataset, on near-threshold visual detection, is the planned extension from hearing to vision (Section~\ref{sec:discussion-limits}).

Sergent et al.\ decided the question on a neural readout instead of on report~\cite{sergent2021}. Twenty participants heard a vowel in noise at six levels, a no-sound level and five signal-to-noise levels. In an active session they identified the vowel and rated its audibility, and in a passive session the same sounds were presented without a task. For each participant, a linear decoder trained on the EEG channels to separate the no-sound level from the highest level reduced each trial to one number in each of 53 successive 30~ms windows, the decoder's \emph{distance}, which is the test's readout. At each window three models of the readout given the level competed by held-out likelihood. In the null model the readout does not depend on the level. The graded model has one Gaussian state whose mean follows a logistic curve of the level, with a spread tied to the mean. The two-state model is a mixture of a low state and a high state whose occupancy follows a logistic curve of the level. The group comparison at each window used the \emph{protected exceedance probability}~\cite{rigoux2014}. It starts from the probability that a model is more frequent in the population than every other compared model. That probability is then averaged with the hypothesis that all models are equally frequent. In the active session the two-state model ranked highest after about 250~ms. For the passive session the article reports the variance profile, the trial-to-trial variance of the readout at each level, and a fit of the two-state model, not the three-model comparison. The article infers passive bifurcation, a split of the passive-session responses into two states, from three observations. The first is a variance burst: a peak of that variance across levels, here at the two highest levels. The second is a correlation with active-session behaviour, and the third is mind-wandering reports. Section~\ref{sec:example-methods} gives our re-implementation of this comparison on the released data, and Section~\ref{sec:example-results} what it reproduced.

\subsection{The LLM Adaptation}\label{sec:example-adaptation}
The LLM side of the example is a study that is designed but held back for a separate report. Its LLM is Qwen3.6-27B, an openly released model with 27 billion parameters and 64 layers, run with each of its weights stored in four bits (\SMdesign{} and \SMpilot). Its procedure is what the audit calibrated. Gurnee et al.\ report, in a preprint, that production LLMs carry a small, sparsely active set of token-linked directions in their middle layers~\cite{gurnee2026}. A \emph{linear lens} is a fitted linear map from a layer's activations to the vocabulary. It decodes these directions to tokens, the LLM's vocabulary units. The preprint reports three further properties of the directions. They are broadcast more widely than other representations. Their set is capacity-limited. And the set is coherent only in an intermediate band of layers. In a further experiment, the input embeddings of two tokens, the vectors with which the LLM represents them, are mixed with a varying weight. The \emph{decoded share} is the position of the activation between the two inputs' own pure endpoints. In early layers the decoded share tracks the weight smoothly. From the onset of the intermediate band, it switches sharply between the two inputs at a threshold value. The theory's authors call the lens space a testable version of their hypothesis. They propose a test with two parts. The first is a threshold nonlinearity at the lens space, against earlier layers whose response rises without a threshold. The second is bimodality across runs, that is, across repeated presentations of the same stimulus exactly at threshold~\cite{dn2026}. A commentary by two authors of the indicator method finds the evidence strong that these directions are privileged. Privileged here means more widely broadcast than other representations. The commentary asks for more evidence of a workspace-like structure~\cite{eleos2026}. The indicator method's own case study had judged transformer LLMs a weak case for the workspace indicators, because they lack recurrence and a workspace distinct from the modules~\cite{butlin2023}. The held LLM study operationalizes the proposed test in three ways. Its dose is a count of clues. Its readout is not restricted to the lens space. And its variability comes from different stimuli in place of repeated runs (\SMdesign). The comparison asks whether one property the theory attributes to the workspace, all-or-none entry, has an analogue in the LLM's readout; in both accounts the workspace broadcasts its content and makes it reportable. Other architectural parallels are left aside.

A reader of Table~\ref{tab:methods} may ask how the methods there would fare on this example. The correspondence methods of Section~\ref{sec:related-comparison}, representational similarity analysis, encoding models, centered kernel alignment and behavioural benchmarks (error consistency), compare two systems' responses to one shared stimulus set, and the example's two sides share none: a vowel in noise on one side, a text packet on the other. The lack of shared stimuli is by design. The CEA scores competing models of each system's own readout and needs no shared items and no map between the systems. The correspondence methods return a similarity between the systems, not a decision on the distribution within one of them. The comparisons that can be made on the same data are made, and Section~\ref{sec:example-results} reports them. On the LLM side, the literature's own tests for two states, the dip test of unimodality, the Bayesian information criterion for a two-component Gaussian mixture and the logistic steepness of the aggregate ignition index, were run on the same simulated datasets as the audited procedure. On the brain side, the same tests and the published test's own rule were scored on the same simulated EEG sets as the matched comparison (Section~\ref{sec:example-methods-matched}).

One concept shows the objects the procedure works with. Take one animal in the bank of animals, foods and vehicles as the target concept. Each concept carries descriptive \emph{clauses}, derived from published lists of the properties that people associate with concepts (feature norms). Each clause is true of its concept and does not name it. A stimulus is a \emph{packet} of eight clause slots. The packet is set inside a \emph{carrier}, one of six fixed text frames. At dose $k$, $k$ of the eight slots hold \emph{clues}, the clauses about the target concept, and the other slots hold background clauses about other concepts. The doses form the set $\mathcal{K} = \{0, 1, 2, 3, 4, 6, 8\}$ of clue counts $k$, so the dose is a count and not a measured scale of evidence. One \emph{trial} is one forward pass, a single evaluation, of the frozen LLM over one packet. The trial's readout is the output of a \emph{decoder}: a linear classifier trained at one layer to separate full-dose from zero-dose packets and then frozen. Let $j$ be the depth index. It runs over layers here and over time windows in the human test, and the two axes are not equivalent. The readout of a packet $x$ is $y_j = r_j(h_j(x))$, where $h_j(x)$ is the residual-stream state at layer $j$ and $r_j$ is the decoder's fixed linear functional. A concept's trials are its packets in every carrier at every dose, with $D$ draws of each, so a concept has $6 \times 7 \times D = 42D$ trials (six carriers, seven doses), 168 at $D = 4$. A fixed checkpoint is a deterministic function of its input, so at a fixed dose and concept every trial-to-trial variation in $y_j$ comes from the stimulus draw: the carrier, the clue positions, the selected clues and the fillers. The object of inference is the conditional law of the readout given the dose and the concept at each depth, and the question is which of two families describes it better out of sample. A \emph{simulated dataset} replaces the LLM's readouts with values drawn from a generator (Section~\ref{sec:audit-loop}). In the earlier draft design used for the audit (Section~\ref{sec:example-methods-calibration}), a dataset holds 64 concepts in eight \emph{concept families}, groups of related concepts with eight concepts in each, and 10{,}752 trials in all.

In the workspace account, the readout observes one channel. In that channel, processors offer candidates in parallel. A capacity-limited stage selects one candidate and broadcasts it, and downstream computation reads the broadcast. With $S$ the state of that stage on a trial, as in Section~\ref{sec:audit-nats}, the theory's claim is that $S$ is binary near threshold. A content's presence is then carried by at most one bit per trial, while the content itself ``can be of variable intensity''~\cite{dn2026}. The graded alternative is that $S$ is continuous, so that the readout's law moves smoothly with the dose through one state. The two families below operationalize these two claims statistically, with the caveat of Section~\ref{sec:audit-nats}. The comparison tests two-state structure within a concept at a fixed dose. A graded decision whose selected member has a steep dose slope and differing concept thresholds remains compatible with all-or-none entry per concept. In the theory the two states arise from fluctuations at a fixed input, whereas here the variability is placed in the input~\cite{dehaene2003,dehaeneChangeux2011,dn2026}. These definitions are those of the LLM-side protocol draft, to be frozen before any confirmatory LLM data are collected, and no confirmatory LLM data exist. The bounded comparison, a revised procedure still in development (Section~\ref{sec:example-methods-bounded}), changes the competing models, the fitting, the statistic, the estimand and the decision rule.

\subsection{The Two Families}\label{sec:example-families}
Both families use a logistic curve in the dose $k$ that rises from 0 to 1, crosses one half at the threshold dose $k_0$ and has steepness $\kappa$: $\mathrm{lg}_{\kappa, k_0}(k) = \bigl(1 + e^{-\kappa (k - k_0)}\bigr)^{-1}$. The graded family $G$ has one state whose location follows the dose. Its base member, inherited from the human comparison, is
\begin{equation}
\begin{aligned}
y \mid k &\sim \mathcal{N}\bigl(\mu(k), \sigma(k)^2\bigr), \\
\mu(k) &= L\bigl[\mathrm{lg}_{\kappa, k_0}(k) - \mathrm{lg}_{\kappa, k_0}(k_{\max})\bigr] + \mu_{\max}, \qquad
\sigma(k) = \lvert a\,\mu(k) + b \rvert,
\end{aligned}
\label{eq:m2b}
\end{equation}
where $\mathcal{N}(\mu, \sigma^2)$ is the normal law with mean $\mu$ and spread $\sigma$, $k_{\max}$ is the highest dose and $\mu_{\max}$ the mean there, $L$ scales the rise to $\mu_{\max}$, and the slope $a$ and intercept $b$ tie the spread to the mean, so that the spread may rise or fall with it. This member is the published graded model. Its mean $\mu(k)$ changes in one direction with the level, so its variance $\sigma(k)^2 = (a\,\mu(k) + b)^2$ can rise, fall or dip in the middle, but cannot peak in the middle. Of the three published models, only the two-state model can therefore fit a variance peak at an intermediate level. Three members widen the family in directions in which a one-state law can be mistaken for two. The first adds a concept random effect on the threshold, $k_0 \to k_0 + \chi_c$ with $\chi_c \sim \mathcal{N}(0, \tau^2)$, where $\tau$ is the spread of thresholds across concepts. The second adds a concept random scale, $\sigma(k) \to \sigma(k)\, e^{v_c}$ with $v_c \sim \mathcal{N}(0, \omega^2)$, where $\omega$ is the spread of log scales across concepts. The third is a skew-normal law $y \mid k \sim \mathrm{SN}\bigl(\xi(k), s_{\mathrm{SN}}(k), \alpha(k)\bigr)$ with location $\xi(k)$ of the form of equation~\eqref{eq:m2b}, scale $s_{\mathrm{SN}}(k) = \lvert a\, \xi(k) + b \rvert$ and shape $\alpha(k) = \alpha_0 + \alpha_1\, \xi(k)$, with intercept $\alpha_0$ and slope $\alpha_1$. Its density is $\frac{2}{s_{\mathrm{SN}}}\,\phi(z)\,\Phi(\alpha z)$ with $z = (y - \xi)/s_{\mathrm{SN}}$, where $\phi$ and $\Phi$ are the standard normal density and cumulative distribution function. The mixture family $X$ has two states and a dose-dependent occupancy:
\begin{equation}
\begin{aligned}
y \mid k &\sim \bigl(1 - A(k)\bigr)\,\mathcal{N}\bigl(\mu_{\mathrm{L}}, \sigma_{\mathrm{L}}^2\bigr) + A(k)\,\mathcal{N}\bigl(\mu_{\mathrm{H}}(k), \sigma_{\mathrm{H}}^2\bigr), \\
A(k) &= \mathrm{lg}_{\kappa_A, k_0}(k) \ \ (k \geq 1), \qquad A(0) = 0, \qquad
\mu_{\mathrm{H}}(k) = \mu_{\mathrm{L}} + e^{\delta_0} + e^{\delta_1}\,\mathrm{lg}_{\kappa_h, k_0}(k),
\end{aligned}
\label{eq:m3}
\end{equation}
where $\mu_{\mathrm{L}}$ and $\sigma_{\mathrm{L}}$ are the low state's mean and spread, $\mu_{\mathrm{H}}(k)$ and $\sigma_{\mathrm{H}}$ the high state's (tied, $\sigma_{\mathrm{H}} = \sigma_{\mathrm{L}}$, except in the member with separate component scales), $A(k)$ is the probability of the high state at dose $k$, $\kappa_A$ and $\kappa_h$ are the steepness of the occupancy and of the high state's rise, and $e^{\delta_0}$ and $e^{\delta_1}$ are the offset and amplitude of that rise. In equation~\eqref{eq:m3} the high state lies above the low state at every dose by construction, and $A(0) = 0$ at the catch level, the dose of zero clues, so that the mixture there is the low state alone. Three members of $X$ keep the high state above the low state, and we call them the \emph{ordered} members. Each adds one freedom to this form. The first adds a concept random effect on the shared threshold. The second has separate component scales in place of the common spread. The third has a free catch-level occupancy $A(0) = \pi_0$, so that spontaneous entry can be read from the no-signal trials~\cite{mashour2020,vanVugt2018,kouider2010}. The inherited two-state model, in which the ordering of the states is not enforced, is kept as the fourth member of $X$ and a comparator, in its published statistical form, with the parameter bounds and scaling of the implementation. The parameters of every member are free per depth and fitted by maximum likelihood. For a member with a concept effect, the likelihood of a concept is the integral over that effect. The families are not nested, by design. Each family adds to its base member the mechanisms, chosen by us, by which a law of one family could imitate the other: heterogeneity across items (concepts here, participants in the human test), a scale that grows with the mean, and skew (the last in $G$ only). The choice follows the finding that a single continuum can yield report distributions previously read as all-or-none~\cite{cohen2023graded}. A mixture's low state need not sit at zero~\cite{overgaard2006,whyteSmith2021}, so the mixture members carry a free low state.

\subsection{The Statistic and the Decision Rule}\label{sec:example-statistic}
For a member $m$ and a held-out concept $c$ with trials $i = 1, \dots, n_c$, the joint held-out likelihood is
\begin{equation}
q_{m, c} = \int \prod_{i = 1}^{n_c} p_m\bigl(y_{ci} \mid k_{ci}, \chi\bigr)\, p_m(\chi)\, \mathrm{d}\chi ,
\label{eq:q}
\end{equation}
where $y_{ci}$ and $k_{ci}$ are the readout and dose of trial $i$ of concept $c$, $p_m$ is the density of member $m$ and $\chi$ its concept effect, with the integral absent where the member has no concept effect. The concepts are split into outer folds, and each outer fold is held out in turn. For each held-out outer fold, every member is refitted on the remaining concepts. The training concepts are split again into inner folds that share no concept, and the member of each family with the best score on these inner folds is selected, $m^\ast_G$ for the graded family and $m^\ast_X$ for the mixture family. A family scores the held-out concept by the likelihood of equation~\eqref{eq:q} of its selected member, $q_{G, c} = q_{m^\ast_G, c}$ and $q_{X, c} = q_{m^\ast_X, c}$. The per-concept statistic is the difference of the two families' held-out log scores per trial,
\begin{equation}
\Delta_c = \frac{\log q_{X, c} - \log q_{G, c}}{n_c} .
\label{eq:delta}
\end{equation}
Equation~\eqref{eq:delta} is in nats per trial, and it is positive where the mixture family predicts the concept's trials better. Conditional on the fitted models, its expectation is equation~\eqref{eq:ce} with $P$ the true joint law of the concept's trial vector, $Q_1$ the graded family's predictive law and $Q_2$ the mixture family's. Because the joint likelihood integrates the concept effect, this is a joint predictive score per trial and not a sum of single-trial terms. The sign and size of $\Delta_c$ are properties of the complete fitted procedure at the design size, because approximation, estimation, family selection and numerical fitting all enter. They are not a theorem about the generating law. The audit loop measures their behaviour under declared generators (Section~\ref{sec:example-methods}).

The band statistic is the mean of $\Delta_c$ over the concepts of a dataset and over the layers of a band,
\begin{equation}
\bar\Delta_{\mathrm{ws}} = \frac{1}{\lvert \mathcal{J}_{\mathrm{ws}} \rvert} \sum_{j \in \mathcal{J}_{\mathrm{ws}}} \frac{1}{\lvert \mathcal{C} \rvert} \sum_{c \in \mathcal{C}} \Delta_{c}^{(j)},
\label{eq:band}
\end{equation}
where $\mathcal{J}_{\mathrm{ws}}$ is the set of layers in the workspace band (layers 23--57; \SMdesign), $\mathcal{C}$ the set of concepts, $\Delta_c^{(j)}$ the statistic of concept $c$ at layer $j$, and $\lvert \cdot \rvert$ the number of elements of a set. The protocol draft takes the name workspace band from the lens study, and the name does not presuppose that the LLM has a workspace. The early and late bands define $\bar\Delta_{\mathrm{early}}$ and $\bar\Delta_{\mathrm{late}}$ alike. The difference $\bar\Delta_{\mathrm{ws}} - \bar\Delta_{\mathrm{early}}$ is the \emph{locality contrast}. It measures how much more the mixture family is favoured in the workspace band than in the early layers. The audit reported here ran at one layer, so its band is that layer. The decision rule reads a 95\% interval for the band mean of equation~\eqref{eq:band}, with the concept as the cluster. An interval above zero is a two-state decision, one below zero a graded decision, and one that includes zero is inconclusive. The interval's target is the estimand of Section~\ref{sec:audit-loop}, $\theta_g = \mathbb{E}_g[\bar\Delta_{\mathrm{ws}}]$, the expectation of the complete procedure's band mean under a generator $g$ at the design sizes. The audit loop measures the interval's bias-eliminated coverage, equation~\eqref{eq:cov}, which is not the coverage of $\theta_g$ itself. These five choices are the five components of the audited procedure: the two families, their fitting with selection nested in the folds, the concept as the unit, $\theta_g$ as the estimand and the interval rule. Figure~\ref{fig:instances} sets them beside the human test's.

\section{Methods of the Example}\label{sec:example-methods}
This section gives the methods of the example: the human data and our re-implementation of the published comparison (Section~\ref{sec:example-methods-human}), the audited procedure, at one layer (Section~\ref{sec:example-methods-procedure}), the calibration by simulation (Section~\ref{sec:example-methods-calibration}), the matched comparison, which gives both families trial-level variation in latency and scale (Section~\ref{sec:example-methods-matched}) and the changes to the procedure that are in development (Section~\ref{sec:example-methods-bounded}). Table~\ref{tab:status} gives each procedure and variant compared in this paper one name, with its statistic, target, sampling unit, design size and validation status, where validated means shown on fresh simulated data to meet the declared error rates, and the text uses only those names. The names in Table~\ref{tab:status} other than the published human comparison are ours.

\begin{table}[!t]
\caption{The procedures and variants compared in this paper, with the statistic, target, sampling unit, design size and validation status of each.\label{tab:status}}
\scriptsize
\setlength{\tabcolsep}{3pt}
\begin{tabularx}{\textwidth}{>{\raggedright\arraybackslash}p{17mm} L >{\raggedright\arraybackslash}p{15mm} >{\raggedright\arraybackslash}p{16mm} >{\raggedright\arraybackslash}p{15mm} L >{\raggedright\arraybackslash}p{12mm}}
\toprule
\textbf{Name} & \textbf{Statistic} & \textbf{Target} & \textbf{Sampling unit} & \textbf{Design size} & \textbf{Validation status} & \textbf{Sections} \\
\midrule
Published human comparison and our re-implementation & Protected exceedance probability of each of three models at each 30~ms window, from participants' cross-validated held-out log-likelihoods & None named; a ranking across windows & Participant & 20 participants, 53 windows, active session & Reproduced against the publisher's source data on the three points this paper uses; single windows differ. Its per-window rule, scored on the matched comparison's simulated sets, called two states in 7 of 50 two-state sets and in 39 of 100 sets resampled from the recorded trials & \ref{sec:example-methods-human}, \ref{sec:example-results} \\
\addlinespace
Matched comparison & Participant mean of the held-out score difference per trial between the two-state and graded members with latency and scale variation, with a 95\% $t$ interval & The mean contrast across a setting's replicate sets & Participant & 20 participants, 16 windows from 255 to 705~ms, five block folds & Calibrated before the real contrast was read: two-state data detected in 49 of 50 T1 sets, none of 100 sets at the margin fell below it, and a shared-covariance defect found and repaired in development. A decision still requires the graded-setting and completion conditions, so the real contrast is reported descriptively. An upper bound below the margin rests on its own conditions, which were met & \ref{sec:example-methods-matched}, \ref{sec:example-results} \\
\addlinespace
Audited procedure & Mean over held-out concepts of $\Delta_c$, the held-out log-score difference per trial of concept $c$, selected mixture member minus selected graded member, in nats per trial & $\theta_g$ & Concept & 64 concepts in eight concept families of eight, $D = 4$, one layer, five outer folds & No false two-state decision in 12{,}000 graded datasets; two-state on 599 of 600 two-state datasets; its bias-eliminated coverage, the share of intervals containing the mean of the estimates, was below 0.90 at six of twelve graded settings, so it cannot support confirmation & \ref{sec:example-methods-procedure}, \ref{sec:example-results} \\
\addlinespace
Ensemble variant & As the audited procedure, with the equal-weight ensemble of each family's joint likelihoods in place of selection & --- & Concept & The audited procedure's fits & Graded on all 12{,}000 graded datasets; a sensitivity check on the same fits, not an independent replication & \ref{sec:example-methods-procedure}, \ref{sec:example-results} \\
\addlinespace
Inherited pair & As the audited procedure, with the two models of the human study scored against each other and no selection & --- & Concept & The audited procedure's fits & 989 two-state decisions and 1{,}147 inconclusive outcomes on 12{,}000 graded datasets, all two-state decisions at settings with item heterogeneity & \ref{sec:example-methods-procedure}, \ref{sec:example-results} \\
\addlinespace
Bounded comparison & Mean over 64 confirmation concepts of each concept's score difference between equal-weight family ensembles, limited to $\pm 0.03$~nat per trial, with a new mixture member whose concepts differ in scale & The median of the study statistic across repetitions of the design & Concept & One layer, three outer folds, the three-family concept design & In development. On saved rows, without the new member, its interval contained the full-bank median (Section~\ref{sec:example-methods-bounded}) in 299 or 300 of 300 assessment datasets (Section~\ref{sec:results-bounded}) per graded setting, with no two-state decision. Calibration and validation on new seeds have not run & \ref{sec:example-methods-bounded}, \ref{sec:example-results} \\
\bottomrule
\end{tabularx}
\end{table}

\subsection{Human EEG: The Data, the Published Models and Our Re-implementation}\label{sec:example-methods-human}
The data are the recordings of the twenty participants released by Sergent et al.\ on the Open Science Framework (project aw3t5)~\cite{sergent2021}, in the active and passive sessions of Section~\ref{sec:example-human}. For each participant and session, a linear logistic decoder is trained on 63 EEG channels to separate the no-sound level from the dataset's highest level. Its held-out signed output, averaged within each of 53 successive 30~ms windows, is the trial's distance, the readout of Section~\ref{sec:example-human}. \SMhuman{} gives the level labels, the channels, the decoder folds and the passive-session sensitivity analysis.

At each window, three models of the distance given the level are fitted. The null model $M_0$ gives the distance one mean and one spread at every level. The graded model $M_2$ is the base member of equation~\eqref{eq:m2b} with the level in place of the dose. The two-state mixture $M_3$ is the inherited two-state model of Section~\ref{sec:example-families}: a common spread for its two states, an occupancy that follows a logistic curve of the level, and a high-state mean that is a scaled logistic curve of the level plus an offset. At the no-sound level the occupancy is zero and the two means coincide. Each model is fitted by maximum likelihood with the Nelder--Mead method at MATLAB's default cap on the number of function evaluations, and the re-implementation's handling of a negative scale is given in \SMhuman. The fitting uses five folds that each hold out whole recording blocks, from blocks 1--20. Four datasets also contain a 21st block, whose trials are always in training and never scored. The cross-validated score of a participant for a model at a window is the mean over the five folds of the fold's summed held-out log-likelihood, with the parameters fitted on every trial outside that fold. The re-implementation keeps the authors' mean over the folds. Section~\ref{sec:example-results} also reports the model selection on the sum over the folds, because the group comparison below depends on the absolute scale of the scores~\cite{rigoux2014}. A log evidence is a prequential sum of predictive scores: each observation is predicted from the observations before it~\cite{kassRaftery1995}. Cross-validated scores are not such a sum. A fold holds on average 179.8 active and 194.2 passive trials. Of the 18{,}054 active and 19{,}526 passive trials retained, 17{,}978 and 19{,}423 are scored at each window.

At each window the participants' scores, in place of the log model evidences that the method assumes, enter random-effects Bayesian model selection~\cite{rigoux2014}. The method estimates the frequency of each model in the population. It returns the \emph{exceedance probability} $\mathrm{xp}_{w,m}$, the posterior probability that model $m$ is more frequent at window $w$ than every other compared model, and the Bayesian omnibus risk $\mathrm{BOR}_w$, the posterior probability that all three models are equally frequent. The protected exceedance probability of Section~\ref{sec:example-human} is then
\begin{equation}
\mathrm{pxp}_{w,m} = (1 - \mathrm{BOR}_w)\,\mathrm{xp}_{w,m} + \frac{\mathrm{BOR}_w}{3},
\label{eq:pxp}
\end{equation}
that is, equation~\eqref{eq:pxp} averages the exceedance probability with the equal-frequency hypothesis, under which each of the three models is most frequent with probability one third. The article's Methods declare a Simes correction across time points~\cite{sergent2021}. The authors' plotting code implements it as a step-up rule at level 0.05 over the 53 windows on $p_w = 1 - \mathrm{pxp}_{w,3}$, the complement of the two-state model's protected exceedance probability. With the $p_w$ sorted in increasing order, $p_{(1)} \le \dots \le p_{(53)}$, the threshold is the largest sorted value $t^\ast$ that satisfies $p_{(i)} \le 0.05\, i / 53$. The marked windows are those with $p_w < t^\ast$, the strict inequality of that code. If no sorted value satisfies the condition, no window is marked. Because the scores are predictive scores and not marginal model evidences, we use the protected exceedance probability as a descriptive convention and not as a $p$-value. The article describes it as remaining above 0.95 for most of the period between 250 and 700~ms. We report three summaries separately: the first window at which it exceeds 0.95, the first run of at least three consecutive windows in which $M_3$ has the highest value, and the Simes-marked set.

The re-implementation keeps three properties of the published design that limit how far its scores are truly held out, listed in Section~\ref{sec:discussion-limits} (ii) and detailed in \SMhuman. It follows the authors' code where their prose and code differ, so it is faithful to their code and does not demonstrate numerical equivalence with the original implementation. The active and passive sessions are analysed separately and never pooled. We add four analyses of our own: a change of the decoder's solver, a restart of the capped fits at six windows, a model selection without $M_0$, and one on the total held-out score. The total held-out score is the sum over the five folds in place of their mean.

\subsection{The Audited Procedure}\label{sec:example-methods-procedure}
The procedure below is the algorithm that the false-positive study and the detection study calibrated. It is specified in the protocol draft, and the changes to it under development are given in Section~\ref{sec:example-methods-bounded}. The graded family $G$ has four members: the base member of equation~\eqref{eq:m2b} and its variants with a concept random threshold, a concept random scale, and skew. The mixture family $X$ also has four: the inherited two-state model in its published statistical form, and three ordered mixtures of equation~\eqref{eq:m3} with a concept random effect, separate component scales, or a free catch-level occupancy (Section~\ref{sec:example-families}). The draft retains every member pending the recovery stage of Section~\ref{sec:example-methods-calibration}, which has not run.

\newcounter{procedure}
\par\medskip\noindent\rule{\linewidth}{0.4pt}\par\nopagebreak
{\small
\noindent\refstepcounter{procedure}\label{proc:layer}\textbf{Procedure~\theprocedure} (one layer). Input: the concepts of a dataset, each with its concept-family label and its trials $(y_{ci}, k_{ci})$, $i = 1, \dots, n_c$; the retained members of $G$ and $X$.
\begin{enumerate}
\item Split the concepts into five concept-disjoint outer folds $\mathcal{T}_1, \dots, \mathcal{T}_5$, stratified by concept family, so that each concept family is spread over the folds; the same folds serve every model and layer.
\item For each outer fold $f$, with the concepts outside it as the training concepts, form four concept-disjoint inner folds of the training concepts, stratified by concept family. For every retained member $m$, fit $m$ on three inner folds, score the fourth by the sum over its concepts of $\log q_{m,c}$ of equation~\eqref{eq:q}, and sum over the four. This inner score $\rho_{m,f}$ is $-\infty$ if any of these fits or scores is not finite.
\item Refit every retained member on the training concepts and score each held-out concept $c \in \mathcal{T}_f$ by $\log q_{m,c}$.
\item In each family select the member with the highest inner score, $m^\ast_G$ and $m^\ast_X$, and compute $\Delta_c$ of equation~\eqref{eq:delta} for $c \in \mathcal{T}_f$ from the selected members' held-out scores. The ensemble variant and the inherited pair, described below, use the other scores of step~3.
\item Form the mean of $\Delta_c$ over the concepts and, across layers, the band mean of equation~\eqref{eq:band}.
\item Draw $B = 2{,}000$ resamples of the concepts with replacement within each concept family, and recompute the mean from the fixed out-of-fold scores for each resample. The interval's endpoints $\ell_r$ and $u_r$ are the 2.5\% and 97.5\% empirical quantiles of the resampled means, and the claimed standard error is their standard deviation.
\item Decide: unavailable if the procedure failed or the interval is not finite; a two-state decision if $\ell_r > 0$; a graded decision if $u_r < 0$; inconclusive otherwise. The first case takes precedence, and an unavailable decision is recorded as a \emph{procedure failure}.
\end{enumerate}
Every fit uses L-BFGS-B, a limited-memory quasi-Newton optimizer with bounds, on analytic gradients and from several starts. The starts are built from data moments and a seeded jitter, with 4 starts in step~2 and 8 in step~3 in the audited configuration. The kept solution is the converged start with the highest finite training likelihood. If no start converges, a recovery chain adds 16 starts, and then 16 more at twice the jitter. If none of those converges, the best finite non-converged start is kept and flagged. A fit with no finite likelihood is unscorable.\par}
\nopagebreak\noindent\rule{\linewidth}{0.4pt}\par\medskip

The estimand of this procedure is the out-of-sample joint log-score advantage per trial of the training-selected mixture member over the training-selected graded member, $\theta_g$ of Section~\ref{sec:example-statistic}. A positive result rejects those graded members, not every graded account. Two comparison variants use the same fits. The \emph{ensemble variant} scores the equal-weight ensemble of each family's joint likelihoods in place of the selected member. The \emph{inherited pair} scores the two published models of Section~\ref{sec:example-methods-human} against each other, with no selection. It is the transferred test with its models carried over unchanged. In the draft the interval of step~6 is the primary interval. If the bias-eliminated coverage of Section~\ref{sec:example-methods-calibration} falls below 0.90 under any retained generator, the draft requires the replacement interval of Section~\ref{sec:example-methods-bounded}, chosen before confirmatory data are collected. The procedure fails, and its decision is unavailable, when in some outer fold every member of a family has an inner score of $-\infty$. It also fails when a retained member cannot be scored on its held-out concepts, because the two variants need every member's scores. The implementation's result flags and its numerical checks are given in \SMaudit.

\subsection{Calibration by Simulation}\label{sec:example-methods-calibration}
A generator is a retained member at declared parameters (Section~\ref{sec:audit-loop}). The completed simulation used the earlier draft design: eight concept families of eight concepts, and $n_c = 42D$ trials per concept from six carriers, seven doses and $D = 4$ draws of each carrier and dose, so 168 trials per concept and 10{,}752 in all. \SMdesign{} describes the revised three-family design (animals, foods and vehicles), at which any replacement of the audited procedure must be validated before confirmation. The simulated readout stands in for the decoder's output at one layer. The LLM, the decoders and the stimuli are not simulated, and the band rule over the 35 layers of the workspace band (\SMdesign) is to be validated in its own stage after the development run, the end-to-end run on real LLM activations of Section~\ref{sec:results-pilot}.

The \emph{false-positive study} drew $R = 1{,}000$ datasets under each of twelve graded generators, 12{,}000 in all. The twelve are the base member of equation~\eqref{eq:m2b} and eleven settings of its three variants: the spread of concept thresholds $\tau$, the spread of concept scales $\omega$ and the skew. Each is varied from a zero setting beside the base member, for the reason given in Section~\ref{sec:audit-loop}. On these datasets the audit measured the false-positive rate of equation~\eqref{eq:fpr}, the mean of the estimates and bias-eliminated coverage of equation~\eqref{eq:cov} and, where no false positive occurred, the bound of equation~\eqref{eq:binom}, which is 0.003 at $R = 1{,}000$. The draft's calibration rules are a false-positive rate of at most $0.05 + 2\sqrt{0.05 \times 0.95 / R}$, which is 0.064 at $R = 1{,}000$, and a bias-eliminated coverage of at least 0.90, at every graded generator. The conventions for failed replicates are given in \SMaudit.

The \emph{detection study} drew 50 datasets under each of twelve two-state generators, 600 in all: the four mixture members, each at three declared gains. The gain is defined per trial. To set an effect size, both high-state offsets of a mixture member, the terms $e^{\delta_0}$ and $e^{\delta_1}$ of equation~\eqref{eq:m3}, are multiplied by a scale $\lambda$. The \emph{gain} is the expected advantage per trial of this scaled member over the graded member with the highest training likelihood, on test concepts drawn independently of the training concepts. The protocol draft states how $\lambda$ is found for each pair of member and gain, the acceptance gate the pair must meet, and the fallback if the search meets a discontinuity. The detection study used the twelve accepted scales, recomputed under its own code and configuration and reproduced exactly before it ran. We report no other gain result. The draft specifies two further stages, a power stage at 1{,}000 datasets per alternative and a recovery stage over a separate grid of 48 generator points, the twelve graded generators and 36 mixture points, which reports the family confusion matrix. Neither has run, and the detection study does not replace the power stage (\SMaudit).

\subsection{The Matched Comparison}\label{sec:example-methods-matched}
A long-established hazard of the human test is trial-to-trial variability in the amplitude and latency of an evoked response, which changes the variance and other statistics of the response at a fixed latency~\cite{truccolo2002}. The published models describe one number per trial, the distance at one window, and say nothing about when the response arrives. If the response arrives at a different moment on each trial, a fixed window sees early responses on some trials and late responses on others, and the variance across trials rises even when the response itself is graded. A two-state mixture can fit such variance better than a graded model without any switching in the brain, and comparing consecutive windows does not control that latency variation. This creates an asymmetry: the two-state mixture can absorb latency variation in this way and the graded model cannot. The \emph{matched comparison} removes that asymmetry. It gives both families the same two sources of trial-to-trial variation, a latency shift and a scale factor, and asks the same question again on the same data. The LLM side has no latency axis.

The data are the decoder distances of the twenty participants, one trial at a time. A trial is now a vector of 16 numbers, the distance averaged within each of the 16 windows between 255 and 705~ms after the sound, the primary span. The comparison scores the 16 numbers of a trial together, which extends the one-number readout of Section~\ref{sec:audit} to several numbers per trial. The six levels are $k = 0, \dots, 5$, with level 0 the no-sound trial, and the scoring folds are the five block folds of the published design. Both families write a trial vector as a baseline, plus a response template multiplied by an amplitude, plus Gaussian noise with a full covariance across the 16 windows. The baseline is the mean of the training no-sound trials. The template $\psi(t)$ is a smooth curve of time $t$, a cubic spline with knots every 60~ms, fitted separately inside each model. On each trial the template is shifted in time by a latency $\eta$ drawn from a normal law with spread $s_\eta$ and cut off at $\pm 150$~ms, and multiplied by a scale factor $\gamma$ drawn from a lognormal law with spread $s_\gamma$ and mean one. The graded family gives every trial at level $k$ the amplitude $\beta(k)$; it is the variable-signal-plus-noise model of~\cite{truccolo2002} with a smooth template. The two-state family gives each trial a state. With probability $A(k)$ the response is present, with amplitude $\beta(k)/A(k)$, and otherwise it is absent. As in equation~\eqref{eq:m3}, $A(k)$ is a logistic curve of the level. If the two families share the amplitude, template and latency law, the division by $A(k)$ gives them the same mean response over trials at each level. These quantities were fitted separately, so the fitted means could differ. Each family has four members, without either variation term, with the latency term only, with the scale term only, and with both: G0 to G3 for the graded family and B0 to B3 for the two-state family. The comparison that answers the question is G3 against B3.

Each model was fitted by maximum likelihood, with the latency and the scale integrated numerically and L-BFGS-B run from several starting points built from the training trials (\SMjitter). An expectation--maximization loop estimated the noise covariance inside each model, and held-out trials were scored by their log density under the fitted model, with that model's own covariance. The score of a model is the held-out log density summed over the five folds and divided by the number of scored trials, in nats per trial. The primary quantity was the participant mean of the score of B3 minus that of G3, with its 95\% $t$ interval. It targets the mean signed predictive advantage across participants, whereas the published protected exceedance probability concerns the proportion of participants best fitted by each model~\cite{rigoux2014}. The decomposition reported in Section~\ref{sec:example-results} uses the same difference for B0 against G0 and for B2 against G2.

Calibration on simulated participants preceded scoring of the matched real-data contrast, and \SMjitter{} gives its full design. Each simulated set drew 20 participant configurations from the participants' fitted models. Four graded settings used fitted graded generators: W1 without latency or scale variation, W2 with their fitted spreads, W3 with an 80~ms latency spread and W4 with a scale spread of 1.0. Five \emph{departures}, O1--O5, were settings that depart from both fitted families, to challenge them. O5 resampled recorded trial vectors within each level and recentred them on the graded generator's marginal means, so its data-generating model was unknown. T1 used the fitted B3 generators, and T2 gave them an 80~ms latency spread and half their state probabilities. A setting's reference contrast is the mean contrast across its replicate sets. The margin $\epsilon$ is one third of the reference contrast under fitted B0 generators, which have no variation terms. E1 and T3 scaled the two-state amplitude so that their reference contrasts were $1.25\epsilon$ and $2\epsilon$. An available set read two-state when its interval lay above zero, graded when it lay below zero, and inconclusive otherwise. Four pass conditions were fixed before validation, with unavailable sets counted against each. The graded-setting condition allowed at most 1 of 100 sets to read two-state under each graded setting and departure. The detection condition required at least 45 of 50 T1 sets to read two-state. The margin condition allowed at most 4 of 100 E1 sets to fall below the margin, that is, to have an upper interval endpoint below $\epsilon$. The completion condition allowed at most 10\% unavailable sets at every attempted setting. A set with fewer than 16 usable participants counted as unavailable. A primary decision required the graded-setting, detection and completion conditions, at least 16 usable real participants, and a finite interval. Real-data latency detection required profile-likelihood lower bounds of the latency spread (the hull construction is in \SMjitter) above 20~ms in at least 10 of 20 participants, and a positive group interval for the score gain from adding latency after scale in that family. The \emph{published per-window rule}, applied to each simulated set, counted a two-state decision at a run of at least three consecutive windows with $M_3$ highest, starting within the primary span.

\subsection{Changes to the Procedure in Development}\label{sec:example-methods-bounded}
Three changes to the audited procedure were considered after the false-positive study. The first is a replacement interval, the \emph{pipeline-refitting bootstrap}, in which every resample reruns fold construction, fitting, family selection and scoring to recompute the statistic (steps 1--5 of Procedure~\ref{proc:layer}). The second is a fixed and richer start recipe for each member, declared and implemented before the runs that calibrate and validate the procedure (\SMrepair). Adoption of an earlier candidate, a start policy in which each fit selects its solution from a declared pool of starts inside the nested procedure, still requires its own paired assessment and complete-procedure validation, including failure rates and computational cost. The third is a calibrated critical value on the point statistic in place of the interval decision of Procedure~\ref{proc:layer}. It would make the procedure a different hypothesis test with its own null calibration. It has since been implemented, with a bounded statistic and a changed target, and screened as development evidence, as described below. The screen used the saved rows, the stored per-dataset outputs of the false-positive and detection studies. A replacement interval has to show a bias-eliminated coverage, against the mean of the estimates $\hat\theta_g$ of a named estimand $\theta_g$, at or above a floor fixed in advance, under the declared graded generators at the design sizes and on seeds not used in its development.

Alternative fixed-score bootstrap intervals, evaluated on the saved rows of the false-positive study, reached a bias-eliminated coverage of 0.90 at no more than nine of the twelve settings. Three development comparisons of the refitting candidate produced no false two-state decision (\SMrepair). They are development comparisons and not a validation of coverage of the unknown expectation. The validation of the refitting interval was staged. Before any coverage was computed, further spending on it was stopped in favour of the less expensive candidate described next, and no coverage result from it is reported here. \SMrepair{} gives the stages, the dated amendments and the measured computational cost.

The less expensive candidate, the \emph{bounded comparison}, removes the bootstrap from the interval. Calibration will repeat the declared training fits and held-out scoring on independent simulated datasets from the frozen generator catalogue, so that fitting variability enters through the distribution of the statistic across datasets. Within each family, the joint predictive densities of the members are averaged with equal weights, so that no member is selected. The mixture family gains a member with a concept random scale, the analogue of the graded member with that effect. Each concept's held-out score difference is limited to the range $-0.03$ to $+0.03$~nat per trial, and the target becomes the median of the study statistic across repetitions of the design. A two-state decision requires both a positive lower interval endpoint and a statistic above a critical value calibrated under the graded generators, both taken as the least favourable over the catalogue. Limiting the scores means that the contrast is no longer a proper scoring rule. The proposed first application uses one layer, three outer folds and the three-family concept design of \SMdesign. Its calibration and validation on new seeds are specified but have not run. They include near-boundary graded data (generators whose expected score difference lies close to zero), graded generators outside the fitted families and two-state data whose concepts differ in scale. \SMbounded{} gives the full specification, its probability bounds and the prerequisites of its calibration. The development screen of Section~\ref{sec:example-results} used a full-bank median: the median over all 1{,}000 saved datasets of a setting. That median includes the setting's 300 assessment rows (Section~\ref{sec:results-bounded}).

\section{Results of the Example}\label{sec:example-results}
This section reports what the instrument delivered on each side of the example (Table~\ref{tab:delivered}), in the order of the audit's questions; the confirmatory comparison of the two systems, which the example was built for, is a further study (Section~\ref{sec:conclusion}).

\begin{table}[!t]
\caption{What the instrument delivered on each side of the example, with its status and the subsection that reports it.\label{tab:delivered}}
\small
\begin{tabularx}{\textwidth}{>{\raggedright\arraybackslash}p{30mm} L >{\raggedright\arraybackslash}p{26mm} >{\raggedright\arraybackslash}p{14mm}}
\toprule
\textbf{Side and data} & \textbf{What was delivered} & \textbf{Status} & \textbf{Section} \\
\midrule
Brain, real EEG & The published decision re-implemented on the open data: first crossing at 315~ms under the authors' convention, resting on 0.0025--0.003~nat per trial; its dependence on solver, rule and score scale measured & Reproduced; single windows differ & \ref{sec:results-human} \\
\addlinespace
Brain, real EEG & The matched comparison, both families with latency and scale variation: $+0.00457$~nat per trial, 95\% interval including zero, upper end below the margin & Descriptive; upper bound's own conditions met & \ref{sec:results-matched} \\
\addlinespace
Brain, simulated sets of known family and resampled sets of unknown family & The matched comparison detected 49 of 50 fitted two-state (T1) sets and none of 100 at the margin fell below it; the published per-window rule called two states in 7 of 50 T1 sets and 39 of 100 resampled sets & Calibrated; graded-setting and completion conditions not met & \ref{sec:results-matched} \\
\addlinespace
LLM, simulated data of known family & The human test's model forms, scored under the audited procedure's sampling unit, interval and decision rule, returned two states on 989 of 12{,}000 graded datasets; the audited procedure did so on none and detected 599 of 600 two-state datasets. Its interval contained the mean of the estimates in fewer than 90\% of datasets at six of twelve graded settings & Calibrated for false positives and detection; interval to be repaired & \ref{sec:results-false}--\ref{sec:results-interval} \\
\addlinespace
LLM, real activations & The procedure ran end to end at one layer on sixteen development concepts with all five components specified & Development run & \ref{sec:results-pilot} \\
\bottomrule
\end{tabularx}
\end{table}

Section~\ref{sec:results-human} reports whether the published human comparison reproduced and how far its result depended on the analysis. Sections~\ref{sec:results-false} and~\ref{sec:results-detection} report how often the audited procedure named the wrong family on simulated data and how often it detected the right one. Section~\ref{sec:results-false} also reports the literature's own two-state tests on the same simulated data of both sides. Section~\ref{sec:results-interval} reports whether the audited procedure's interval contained the mean of the estimates and whether the standard error claimed under its sampling unit matched the replicate spread. Section~\ref{sec:results-matched} reports the matched comparison on the human data, Section~\ref{sec:results-bounded} the development screens of the bounded comparison, and Section~\ref{sec:results-pilot} a development run of the procedure on an LLM.

\subsection{The Human Comparison, Reproduced}\label{sec:results-human}
\begin{figure}[!t]
\centering
\includegraphics[width=\textwidth]{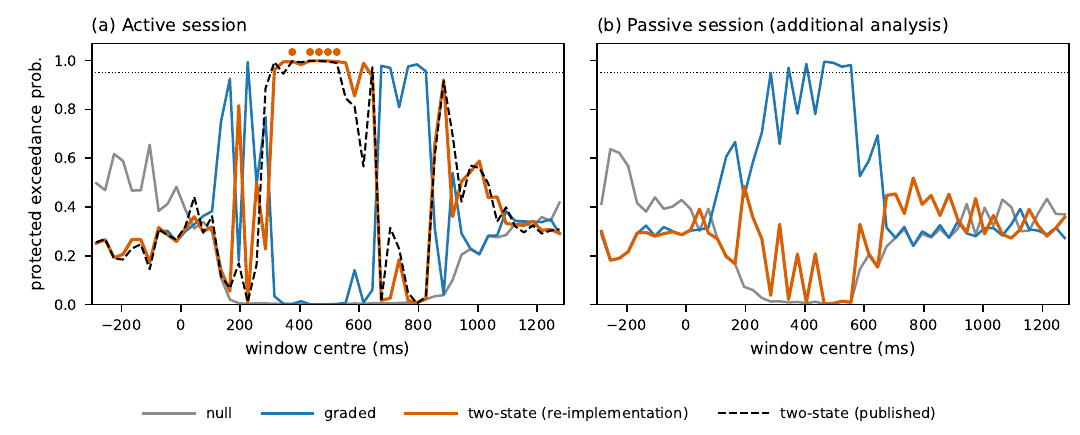}
\caption{The re-implementation's group comparison on the open EEG data of twenty participants, the protected exceedance probability of equation~\eqref{eq:pxp} of the null, graded and two-state models at each 30~ms window under the authors' score convention (solid), for (a) the active session, with the published two-state curve from the publisher's source data (dashed), the 0.95 level (dotted) and the re-implementation's Simes-marked set for the two-state model (dots), and (b) the passive session, an additional analysis not reported in the published study.}
\label{fig:human}
\end{figure}

The re-implementation was run on all twenty participants in both sessions, with the trial counts of Section~\ref{sec:example-methods-human}. We compared its active curve with the publisher's numerical source data, not with the article's prose. Table~\ref{tab:human} and Figure~\ref{fig:human} summarize the comparison. In the active session, the two-state model's protected exceedance probability first exceeded 0.95 at the window centred on 315~ms, both in the published curve and in the re-implementation, under the authors' score convention, the mean over the five folds of Section~\ref{sec:example-methods-human}. The re-implementation also reproduced the broad temporal sequence of the published curve. The null model ranked highest through the baseline, followed by the graded model in an early period up to about 250~ms. The two-state model ranked highest through a late plateau, and the graded model again ranked highest from 675~ms. That agreement is not numerical identity. The first run of at least three consecutive windows in which the two-state model ranked highest began one window later in the re-implementation than in the published curve. The two curves also differed at 555 and 615~ms, and in the published curve the windows above 0.95 were not continuous from 315~ms. These differences enter no result reported here. The simulation audit uses no human data, the matched comparison refits each participant's trials, and the human reference for the LLM study is the active session's two-state preference at the 435--495~ms plateau, which held under every variant below. A crossing of 0.95 is also not a decision corrected for the number of windows tested. In the re-implementation, the Simes-marked set of Section~\ref{sec:example-methods-human} was 375 and 435--525~ms. Applied to one minus the protected exceedance probability, which is not a null $p$-value, that rule controls no error rate, and we report it as the inherited graphical convention for reading the curve across windows.

\begin{table}[!t]
\caption{The active-session group comparison of the published curve, from the publisher's source data, with our re-implementation, under the authors' score convention except in the last two rows, where pxp is the protected exceedance probability of the two-state model, equation~\eqref{eq:pxp}, and times are window centres in ms.\label{tab:human}}
\small
\begin{tabularx}{\textwidth}{L >{\raggedright\arraybackslash}p{36mm} >{\raggedright\arraybackslash}p{36mm}}
\toprule
\textbf{Summary} & \textbf{Published curve (source data)} & \textbf{Re-implementation (ours)} \\
\midrule
Baseline, $-285$ to $+15$~ms & null model ranks highest & null model ranks highest \\
First window with pxp $> 0.95$ & 315 & 315 \\
pxp at 255 / 285 / 315 & 0.164 / 0.886 / 0.994 & 0.500 / 0.227 / 0.963 \\
Start of the first run of $\geq 3$ consecutive windows in which the two-state model has the highest pxp, at any value & 285 & 315 \\
Every window with pxp $> 0.95$, consecutive or not & 315, 375, 405, 435--525, 645 & 315--555, 615 \\
Simes-marked set & not compared & 375, 435--525 \\
pxp at 555 / 615 & 0.846 / 0.565 & 0.991 / 0.989 \\
First window back to the graded model & 675 & 675 \\
\midrule
Correlation of the two curves, 53 windows & \multicolumn{2}{l}{$r = 0.88$} \\
Mean absolute difference, 315--645~ms & \multicolumn{2}{l}{0.063} \\
\midrule
First window with pxp $> 0.95$, total held-out score & not available & 375 \\
Simes-marked set, total held-out score & not available & none \\
\bottomrule
\end{tabularx}
\end{table}

The group result rests on a modest predictive advantage. At the plateau windows, 435--495~ms, the two-state model's cross-validated score exceeded the graded model's in 12 to 14 of the 20 participants. Its mean advantage was 0.43--0.53~nat per participant per held-out fold. A fold holds about 180 scored trials, so this is 0.0025--0.003~nat per trial, with $t \approx 2.5$--$3.0$ across participants. At the same windows, the graded model's mean advantage over the null model was 13--18~nat per fold. With the null model removed from the selection, the two-state model's protected exceedance probability against the graded model alone was 0.62 at 315~ms and 0.81--0.86 at 435--495~ms, against 0.96 and 0.999 with all three models in the selection (\SMhuman). Because the null model fits clearly worse, the Bayesian omnibus risk is near zero, 0.001--0.010 at these windows. The three-model value then approaches the unprotected exceedance probability, which overstates evidence~\cite{rigoux2014}. Including the null model therefore changes the group ranking across participants. It does not change the pairwise predictive-score difference between the two-state and graded models.

The fits used the authors' Nelder--Mead optimizer with its default evaluation cap, and most graded and two-state fits stopped at that cap. Two diagnostics tested how much of the result depended on the fitting. The first replaced the solver of the decoder that produces the single-trial distances. This changed the single-trial distances by less than 0.2\% of their spread, yet it moved the edges of the result: the two-state model then ranked highest from 285~ms, and the late graded period started at 735~ms. The second diagnostic restarted every capped graded and two-state fit at 255, 285, 315, 435, 465 and 495~ms, with 14{,}000 further evaluations. Under the authors' score convention, the 315~ms crossing and the 435--495~ms plateau stayed in place, and the 255 and 285~ms windows moved in both directions. About 30\% of the two-state fits in this restart still stopped at a cap. The windows from 555 to 645~ms were not restarted, so their sensitivity is known from the solver comparison only. The reproduction therefore holds at the tested windows and under the tested optimizer changes, and it is not a converged fit at every window. A third diagnostic changed the scale of the scores, on which the random-effects selection depends~\cite{rigoux2014}. On the total held-out score, in which every scored trial counts once, the two-state model's protected exceedance probability was 0.900 at 315~ms and 0.946 at 345~ms. It first exceeded 0.95 at 375~ms, fell to 0.64 at 405~ms, and was 0.998, 0.990 and 0.995 at 435, 465 and 495~ms, respectively. The first run of three windows in which the two-state model ranked highest still began at 315~ms, and the graded model again ranked highest from 675~ms. No window passed the Simes convention on this scale: no sorted value of one minus the protected exceedance probability was at or below its step-up threshold $0.05\,i/53$ (the smallest, 0.0019, already exceeds $0.05/53$). These shifts are far larger than the Monte Carlo error of the exceedance probabilities. The published curve cannot be recomputed on this scale, because the source data give only its protected exceedance probabilities. The 315~ms crossing therefore holds under the authors' score convention and not on the total held-out score, and the 435--495~ms plateau holds under both.

For the passive session, the article reports no three-model comparison, so ours is an additional analysis under the re-implementation's readout. Under it the graded model ranked highest at every window from 225 to 645~ms, but no window passed the Simes correction for any model (Figure~\ref{fig:human}b). This ranking is a property of the fitted statistical models under one readout, not a demonstration of graded physiology or of the absence of state switching, and model-recovery simulations for this likelihood family have not been run. The two-state model's held-out advantage over the graded model, averaged over 435--495~ms, was larger in the active than in the passive session for 18 of the 20 participants, with a mean paired difference of 0.00375~nat per scored trial and a descriptive paired $t(19) = 4.2$. The windows were chosen from the active-session result and the two sessions use separately trained decoders, so this is a post hoc description, not a confirmatory comparison of the sessions or evidence of a difference in the underlying process. A decision-stage account, in which a criterion is applied to a graded signal~\cite{lau2008,lauRosenthal2011}, would also yield a two-state preference where the task requires a decision and not where it requires none. The active session was designated the human reference for the proposed LLM study after these results were known. \SMhuman{} gives the details, their limits, and the full summary of what the human analyses found.

\subsection{False Positives}\label{sec:results-false}
The false-positive study of Section~\ref{sec:example-methods-calibration} ran Procedure~\ref{proc:layer} on $R = 1{,}000$ simulated datasets under each of the twelve graded generators at $D = 4$. Every retained fit met the recorded convergence criterion, and no fit failure or procedure failure was recorded. The run evaluated the procedure as specified in the protocol draft when the run was launched, on 12~September~2026: four starts for inner selection, eight for refits, and the fixed-score interval of Procedure~\ref{proc:layer}, which resamples concepts while keeping each concept's out-of-fold score fixed. The result is conditional on that start budget, because under-optimized fits can bias either family's held-out score~\cite{wilsonCollins2019}. The twelve generators are the base graded member (M2B), the threshold random effect (M2H) at $\tau \in \{0, 0.5, 1, 2\}$, the skew member (M2K) at three shape settings, $\alpha_0 = 0$, 1 and 3 with $\alpha_1 = 0$, written $\alpha = 0$, 1 and 3, and the concept random scale (M2S) at $\omega \in \{0, 0.5, 1, 2\}$. \SMaudit{} plots the false-positive rate and the bias-eliminated coverage of equations~\eqref{eq:fpr} and~\eqref{eq:cov}, and the replicate spread of the statistic beside the claimed standard errors, at every setting.

The procedure made no false two-state decision at any of the twelve settings: $\hat F_g = 0/1{,}000$ per setting. The exact one-sided 95\% upper bound of equation~\eqref{eq:binom} is then $F^{+} = 0.003$ per setting. This bound is not simultaneous across the settings, and it does not extend to generators that were not tested. All 12{,}000 intervals lay below zero. The mean of the estimates $\hat\theta_g$ was negative throughout the tested grid, from $-0.015$~nat per trial at the base generator to $-5.3$ under the widest concept random scale (the mechanism is in Section~\ref{sec:results-interval}). The negative sign of $\hat\theta_g$ is an empirical property of the procedure at these settings.

Two further comparison variants were scored on the same fits of every dataset (Section~\ref{sec:example-methods-procedure}): the ensemble variant and the inherited pair, the two models carried over from the human study and scored against each other. All three are decided by the same fixed-score concept-cluster interval and the same rule on its sign. A comparison between them therefore holds the sampling unit, interval and decision rule fixed. Only the competing models differ, together with the selection among members that they require. Of the published analysis, just the two model forms are carried over. Its sampling unit, fitting, group statistic and temporal decision are all different (\SMaudit), so nothing here re-runs the human procedure and nothing here calibrates it. Table~\ref{tab:sens} gives all twelve settings, from a post hoc analysis of the saved rows (Section~\ref{sec:example-methods-bounded}) with no refitting. The audited procedure and the ensemble variant returned graded in all 12{,}000 datasets. The inherited pair returned 989 two-state decisions and 1{,}147 inconclusive outcomes. Its two-state decisions occurred at settings where the generator included item heterogeneity that both of its models omit: 651 in the 1{,}000 datasets at $\tau = 2$, where the pair returned no graded decision at all, 337 at $\omega = 2$ and one at $\omega = 1$. There were none at the base generator, none at any of the three skew settings and none at the zero-heterogeneity settings. The audited graded family includes that heterogeneity, so the contrast shows the cost of omitting it from the competing models, not an advantage on generators outside both audited families. A two-state decision under a declared graded generator is false with respect to that data-generating model. It is not a type-I error against a null of zero expected score difference, because this pair is misspecified and its predictive preference between two wrong models can be genuinely positive. The set of twelve settings is our own design, so 989 of 12{,}000 is not a false-positive rate for the pair in general. The audited procedure and the ensemble variant are computed from the same fits on the same datasets, so their agreement is a sensitivity check and not an independent replication. The comparison does show that the zero-false-positive result depends on the expanded families, because the inherited pair alone did not produce it under the same rule, and the ensemble variant, which uses those families without selection, produced none on the same fits.

\begin{table}[!t]
\caption{Decisions of the three comparison variants on the same 12{,}000 graded datasets ($D = 4$, one layer) by graded setting, as graded (G), two-state (T) and inconclusive (I), with the bias-eliminated coverage of the audited procedure's interval, the share of its intervals that contain the mean of the estimates over that setting's replicates, in the last column.\label{tab:sens}}
\footnotesize
\begin{tabularx}{\textwidth}{L >{\raggedleft\arraybackslash}p{11mm} >{\raggedleft\arraybackslash}p{16mm} >{\raggedleft\arraybackslash}p{16mm} >{\raggedleft\arraybackslash}p{24mm} >{\raggedleft\arraybackslash}p{17mm}}
\toprule
\textbf{Graded setting} & \textbf{$R$} & \textbf{Audited G/T/I} & \textbf{Ensemble G/T/I} & \textbf{Inherited pair G/T/I} & \textbf{Coverage} \\
\midrule
Base (M2B) & 1{,}000 & 1000/0/0 & 1000/0/0 & 1000/0/0 & 0.935 \\
Threshold effect, $\tau = 0$ & 1{,}000 & 1000/0/0 & 1000/0/0 & 1000/0/0 & 0.918 \\
Threshold effect, $\tau = 0.5$ & 1{,}000 & 1000/0/0 & 1000/0/0 & 1000/0/0 & 0.819 \\
Threshold effect, $\tau = 1$ & 1{,}000 & 1000/0/0 & 1000/0/0 & 975/0/25 & 0.926 \\
Threshold effect, $\tau = 2$ & 1{,}000 & 1000/0/0 & 1000/0/0 & 0/651/349 & 0.894 \\
Skew, $\alpha = 0$ & 1{,}000 & 1000/0/0 & 1000/0/0 & 1000/0/0 & 0.915 \\
Skew, $\alpha = 1$ & 1{,}000 & 1000/0/0 & 1000/0/0 & 1000/0/0 & 0.885 \\
Skew, $\alpha = 3$ & 1{,}000 & 1000/0/0 & 1000/0/0 & 1000/0/0 & 0.928 \\
Concept scale, $\omega = 0$ & 1{,}000 & 1000/0/0 & 1000/0/0 & 1000/0/0 & 0.917 \\
Concept scale, $\omega = 0.5$ & 1{,}000 & 1000/0/0 & 1000/0/0 & 966/0/34 & 0.848 \\
Concept scale, $\omega = 1$ & 1{,}000 & 1000/0/0 & 1000/0/0 & 846/1/153 & 0.780 \\
Concept scale, $\omega = 2$ & 1{,}000 & 1000/0/0 & 1000/0/0 & 77/337/586 & 0.216 \\
\midrule
All settings & 12{,}000 & 12000/0/0 & 12000/0/0 & 9864/989/1147 & --- \\
\bottomrule
\end{tabularx}
\end{table}

Three properties of these twelve settings limit what the absence of false positives establishes. Each makes false positives less likely than harder generators would. First, every graded setting is a retained member of the family the procedure fits, at declared parameters. The audit therefore measured behaviour under well-specification, that is, with the generator among the fitted models. The settings are adversarial inside the hypothesis space and not outside it. The audit does not address a graded process that the family cannot represent. Second, the graded family has a member with a concept random scale and the mixture family has none, so at the $\omega$ settings an unmodelled spread of scales penalises every mixture member. Third, no interval of the audited procedure in the 12{,}000 came near the decision boundary. The rule calls two states when the interval's lower endpoint exceeds zero, and the largest upper endpoint of its intervals at any of the twelve settings was $-0.0067$~nat per trial. Depending on the setting, the median interval's distance below zero was between 1.2 and 4.8 times its own width. These runs do not assess error control near zero expected score difference. The zero count describes the declared, well-specified settings, all far from the decision boundary, and it does not establish specificity beyond them.

The same-data comparisons of both sides, named in Section~\ref{sec:example-adaptation}, are as follows. On the LLM side, three summaries from the literature were computed post hoc on regenerated versions of the false-positive and detection studies' datasets, with their rules fixed before the analysis ran (\SMalttests). Hartigan's \emph{dip test}~\cite{hartigan1985}, a test of unimodality that measures the largest distance between the empirical distribution function and the closest unimodal one, was applied at dose levels 2, 3 and 4 with a Bonferroni rule over the three levels. It rejected in none of the 12{,}000 graded datasets and in 18 of the 600 two-state datasets of Section~\ref{sec:results-detection}. Two- and one-component Gaussian mixtures were compared by the \emph{Bayesian information criterion} (BIC), a log-likelihood penalized by the number of parameters and the sample size~\cite{kassRaftery1995,schwarz1978}. Two components were selected when the BIC of the one-component mixture exceeded that of the two-component mixture by more than 10 at any of the three levels, with no multiplicity adjustment. This rule selected two components in 5{,}577 graded datasets, at six settings with item heterogeneity or skew, and in 412 two-state datasets. Rahbar's logistic steepness~\cite{rahbar2026} was fitted to the mean readout against dose, not along the depth axis on which its source fitted it. Its medians ran from 0.68 to 1.67 under graded settings and from 1.36 to 1.61 under two-state settings, and its source defines no decision threshold. These outputs describe the tested generators and do not reassess the methods' published results. On the brain side, the same tests were run on the matched comparison's simulated sets, regenerated on their frozen seeds, with every rule declared before the first computation and a check, before any new number, that the regenerated sets reproduce the saved record (Table~\ref{tab:brainside}; \SMalttests). The distance at each window and level was pooled over the set's twenty participants after each participant's no-sound mean was subtracted, the near-threshold levels were 2 to 4, and a two-state decision required three consecutive windows of the sixteen, as in the published reading. The dip test rejected unimodality on no set of any setting, graded or two-state. The fitted two-state generators place their two components between about 1.2 and 1.5 noise standard deviations apart at the template's peak, and a mixture that close is unimodal. The comparison of Gaussian mixtures by BIC preferred two components on most sets of every setting on the pooled distances, from 77 of 100 under W3 to all 100 under W2, W4 and O1 to O4, and including 90 of 100 W1 sets, which carry no trial-to-trial variation. Standardizing each participant's distances by its own no-sound spread removed every two-state decision under W1 and W3, so those decisions came from pooling twenty participants with different noise scales. The two-state decisions persisted wherever the generator carried the lognormal scale term. On the two-state generators as fitted, the standardized reading was 0 of 50 under T1 and 5 of 100 under E1, whose components are too close for a one-dimensional two-Gaussian fit, against 50 of 50 under T2, whose halved occupancy doubles the separation. The published study's own signature is the \emph{variance burst}: an interior maximum of the variance across levels that exceeds both ends by a declared margin of 10\%. The variance burst appeared on at most 2 of 100 sets under the graded generators with or without trial variation, and on at most 3 of 100 under the two-state generators as fitted. It was present on all 100 O5 sets, which resample the recorded within-level distributions, and on the real recordings at every window. The per-window rule and the matched comparison are restated beside these counts, and Section~\ref{sec:results-matched} reports them. No criterion for ranking the procedures was fixed before the runs. We read the counts afterwards on the two-state question alone. On the brain side, the matched comparison made no two-state decision under the graded settings and the departures O1 to O4, and it detected two-state data in 49 of 50 T1 sets. On the LLM side, the audited procedure made no two-state decision on graded data and detected 599 of 600 two-state datasets. No other procedure combined no two-state decision on graded data with detection at these rates. The dip test on the LLM side also made no two-state decision on graded data, but it detected 18 of the 600 two-state datasets. None of these outputs says that the published test is false.

\begin{table}[!t]
\caption{The literature's two-state tests, the published per-window rule and the matched comparison on the matched comparison's simulated EEG sets, as sets reading two-state under the declared rules (distances pooled over the twenty participants, levels 2--4, a two-state decision at three consecutive windows), with the exact one-sided 95\% bound in parentheses (upper on the graded settings and departures, lower on T1, T2 and E1), and the real recordings as one descriptive set.\label{tab:brainside}}
\footnotesize
\setlength{\tabcolsep}{3pt}
\begin{tabularx}{\textwidth}{L >{\raggedleft\arraybackslash}p{7mm} >{\raggedleft\arraybackslash}p{15mm} >{\raggedleft\arraybackslash}p{17mm} >{\raggedleft\arraybackslash}p{17mm} >{\raggedleft\arraybackslash}p{19mm} >{\raggedleft\arraybackslash}p{24mm}}
\toprule
\textbf{Setting} & \textbf{Sets} & \textbf{Dip test} & \textbf{Mixture BIC $> 10$} & \textbf{Variance burst} & \textbf{Per-window rule} & \textbf{Matched comparison} \\
\midrule
W1, graded, no trial variation & 100 & 0 (0.030) & 90 (0.945) & 0 (0.030) & 1 (0.047) & 0 (0.030) \\
W2, graded, fitted spreads & 100 & 0 (0.030) & 100 (1.000) & 0 (0.030) & 0 (0.030) & 0 (0.030) \\
W3, graded, 80~ms latency & 100 & 0 (0.030) & 77 (0.837) & 0 (0.030) & 0 (0.030) & 0 (0.030) \\
W4, graded, scale 1.0 & 100 & 0 (0.030) & 100 (1.000) & 2 (0.062) & 0 (0.030) & 0 (0.030); 4 unavailable \\
O1, time stretch & 100 & 0 (0.030) & 100 (1.000) & 0 (0.030) & 0 (0.030) & 0 (0.030) \\
O2, two template components & 100 & 0 (0.030) & 100 (1.000) & 0 (0.030) & 0 (0.030) & 0 (0.030) \\
O3, level-dependent latency & 100 & 0 (0.030) & 100 (1.000) & 0 (0.030) & 0 (0.030) & 0 (0.030) \\
O4, correlated scale and latency & 100 & 0 (0.030) & 100 (1.000) & 0 (0.030) & 0 (0.030) & 0 (0.030) \\
O5, resampled recorded trials & 100 & 0 (0.030) & 98 (0.996) & 100 (1.000) & 39 (0.477) & 12 (0.187) \\
T1, two-state, fitted & 50 & 0 (0.000) & 44 (0.777) & 0 (0.000) & 7 (0.068) & 49 (0.909) \\
T2, two-state, 80~ms, occupancy halved & 50 & 0 (0.000) & 50 (0.942) & 0 (0.000) & not run & 50 unavailable \\
E1, two-state at the margin & 100 & 0 (0.000) & 92 (0.860) & 3 (0.008) & not run & 99 (0.953) \\
\midrule
Real recordings, active session (descriptive) & 1 & no & yes & yes & --- & --- \\
\bottomrule
\end{tabularx}
\end{table}

\subsection{Detection}\label{sec:results-detection}
The false-positive study's 12{,}000 datasets are all graded, so it cannot measure detection. The detection study measured detection with the identical procedure, interval and decision rule, on datasets generated with a two-state mixture, reported beside the false-positive study and never pooled with it (Section~\ref{sec:audit-loop}). Four mixture members were run at each of three nominal per-trial gains, 0.003, 0.01 and 0.03~nat, with $R = 50$ datasets at each of the twelve settings, 600 in all. M3 denotes the inherited two-state model, and M3H, M3V and M3L the ordered mixtures with a concept random effect on the shared threshold, with separate component scales and with a free occupancy at the catch level, respectively. The run used $D = 4$, the same single layer, 64 concepts per dataset in eight families of eight, four starts for inner selection, and the same fixed-score concept-cluster interval at $B = 2{,}000$. Before the run was launched, the twelve accepted gain scales were recomputed under its code and configuration, and they reproduced the accepted values exactly. The protocol draft sets 0.01~nat as the target, and the other two gains were added to show where detection emerges. Every generator carries a mixture, so a two-state decision is correct with respect to the data-generating model. All 600 datasets were attempted and completed. No dataset recorded a fit failure or a procedure failure, and no inner fit was non-finite, so each detection rate below has all 50 attempted datasets as its denominator.

\begin{table}[!t]
\caption{Detection of the declared two-state alternatives by the audited procedure of Table~\ref{tab:sens}, as two-state decisions over datasets attempted for each of four mixture members at three nominal gains ($D = 4$, one layer, $R = 50$ per setting, 600 datasets), with the exact one-sided 95\% lower bound on the detection probability in parentheses.\label{tab:sens600}}
\small
\begin{tabularx}{\textwidth}{L >{\raggedleft\arraybackslash}p{26mm} >{\raggedleft\arraybackslash}p{26mm} >{\raggedleft\arraybackslash}p{26mm}}
\toprule
\textbf{Mixture member} & \textbf{0.003~nat} & \textbf{0.01~nat (primary)} & \textbf{0.03~nat} \\
\midrule
M3 & 50/50 (0.942) & 50/50 (0.942) & 50/50 (0.942) \\
M3H, $\tau = 0.5$ & 50/50 (0.942) & 50/50 (0.942) & 50/50 (0.942) \\
M3V & 50/50 (0.942) & 50/50 (0.942) & 50/50 (0.942) \\
M3L, $\pi_0 = 0.05$ & 49/50 (0.909) & 50/50 (0.942) & 50/50 (0.942) \\
\bottomrule
\end{tabularx}
\end{table}

Eleven of the twelve settings returned two-state on all fifty datasets (Table~\ref{tab:sens600}). The twelfth was the member M3L at the weakest gain, whose generator sets the catch-level occupancy to $\pi_0 = 0.05$: it returned two-state on 49 datasets and an inconclusive interval on the remaining one. No dataset at any setting returned graded. The worst exact one-sided 95\% lower bound on the detection probability at a single setting was 0.909, and under a twelve-way Bonferroni correction, which makes all twelve bounds hold at once, the worst was 0.857. The draft requires a detection probability of at least 0.8 at 0.01~nat under every mixture member, and the exact lower bounds exceeded that level, both setting by setting and under the Bonferroni correction. These are exact bounds computed from the observed counts, and they do not certify the draft's power stage, which specifies 1{,}000 datasets per alternative. Every setting was at the ceiling or one dataset below it, so the three gains were indistinguishable in outcome. The study set a lower limit on the detection rate and did not locate the gain at which detection begins to fail, so no gain-response curve can be drawn from these results. The one departure from the ceiling was a single dataset, at the smallest gain of M3L, where failure would be expected first.

These are detection rates under the fixed-score rule of Section~\ref{sec:results-false}. They are not size-controlled power, that is, the power of a test whose false-positive rate is known to hold at its nominal level. An interval that is too narrow calls two states more often when the scores are positive, and these rows leave the coverage shortfall of Section~\ref{sec:results-interval} untouched. A high detection rate is therefore consistent with that shortfall and does not repair it. The rates also describe only these twelve declared settings. The alternatives were chosen to be detectable at declared effect sizes, and the results do not assess unmodelled mixtures or the boundary case of zero expected score difference. In particular, none of the four mixture generators gives its concepts different residual scales, a variation that the graded family can absorb and the mixture family cannot, so detection of two-state concepts with different scales still requires independent validation. The reading of Table~\ref{tab:sens600} was blinded. It was written from a table in which each setting carried a coded identifier, and it was committed to the repository before the generator labels were attached (\SMblind).

\subsection{The Interval and the Sampling Unit}\label{sec:results-interval}
The draft requires the bias-eliminated coverage $\hat C_g$ of equation~\eqref{eq:cov} to be at least 0.90 under every graded generator. The bias-eliminated coverage fell below that floor in six of the twelve settings (Table~\ref{tab:sens}). Under the threshold random effect it was 0.819 and 0.894 at $\tau = 0.5$ and 2, respectively. Under skew at $\alpha = 1$ it was 0.885. Under the concept random scale it was 0.848, 0.780 and 0.216 at $\omega = 0.5$, 1 and 2, respectively. Skew at $\alpha = 1$ was among the six, so the shortfall was not confined to concept heterogeneity. Because the reference $\hat\theta_g$ is estimated from the same replicates, these are bias-eliminated coverages against an estimated reference, not coverage probabilities (Section~\ref{sec:audit-loop}), and the plug-in Monte Carlo standard error does not represent that dependence. The six shortfalls differed in size relative to that error. Each setting's plug-in standard error is about 0.010 near the floor at $R = 1{,}000$. Measured in that error, the two smaller shortfalls, 0.894 at $\tau = 2$ and 0.885 at $\alpha = 1$, lay about 0.6 and 1.5 errors below 0.90, respectively, and the other four lay between 4.6 and 53 errors below it. A saved-row bootstrap included uncertainty in the common reference: it resampled complete replicate rows and re-estimated the reference in every resample, with 10{,}000 resamples per setting. Under the threshold random effect at $\tau = 0.5$ and the concept random scale at $\omega = 0.5$, 1 and 2, the bootstrap interval for the bias-eliminated coverage lay wholly below 0.90, and no resample reached the floor. At $\tau = 2$ the interval was $[0.878, 0.916]$ and at $\alpha = 1$ it was $[0.865, 0.904]$, so these two shortfalls were not separated from the floor. The declared rule applies to the six observed rates. The bootstrap gives an approximate uncertainty for this empirical coverage statistic, not a bound on coverage of the unknown expectation, and it is distinct from the pipeline-refitting bootstrap of Section~\ref{sec:example-methods-bounded}, which re-runs the fits. Under the draft's rule, a bias-eliminated coverage below the floor under any retained generator requires a replacement interval, validated on new seeds, before any confirmatory inference. An interval screened offline on these same rows would be method development, not validation.

At $\omega = 2$ the reference is itself unstable. An independent bank of 1{,}000 datasets under the same procedure gave a mean of $-6.87$ (standard error 2.68) but a median of $-2.07$, with every estimate below zero. Removing the single most extreme dataset moved the mean to $-4.22$, which illustrates the influence of one dataset and is not a proposed trimmed estimand. At this setting, the frequency with which intervals contain a candidate value for the reference, whether the bank's mean or median or the mean of the procedure's own estimates, does not establish coverage of the unknown expectation that the interval targets (\SMomega).

The sampling unit was checked against replication. The band point estimate of each dataset is the mean of its 64 held-out per-concept scores. The standard error claimed by the concept-cluster interval was close to the concept-level quantity $\mathrm{sd}(\Delta_c)/\sqrt{64}$, the standard deviation of the 64 per-concept scores divided by $\sqrt{64}$, and smaller by a near-constant factor. At the base generator the median recomputed value was 0.00157 against a median claimed value of 0.00147. The two are different estimators: the interval resamples whole concepts with replacement within each of the eight family strata, while the recomputation pools all 64 concepts and divides by 63. Across the twelve settings the claimed value had a median of 0.938 of the recomputed one, between 0.936 and 0.941, against the $\sqrt{7/8} = 0.935$ that stratified resampling predicts. For a mean of fixed independent scores, resampling within families of eight gives a variance of 7/8 of the usual unbiased stratified estimate. This finite-sample resampling factor therefore narrows the claimed uncertainty. Representational similarity analysis resamples its stimulus conditions so that its inference generalizes to a population of stimuli~\cite{kriegeskorte2008}. The narrowing that duplicated units cause in such resampling is also known there. The factor does not assess the covariance that fitting and selection introduce. On the saved false-positive-study rows, widening each interval about its point by $\sqrt{8/7}$ raised bias-eliminated coverage above 0.90 at $\tau = 2$ (0.921) and $\alpha = 1$ (0.902), while the other four settings below the floor stayed below it, at 0.849 ($\tau = 0.5$), 0.877 ($\omega = 0.5$), 0.806 ($\omega = 1$) and 0.223 ($\omega = 2$).

Whether the interval claims the uncertainty the statistic has can be checked against the spread the statistic shows when the generator is run again. \SMaudit{} sets the claimed uncertainty beside the replicate standard deviation of the point estimate over each setting's 1{,}000 datasets, aggregated post hoc from the saved rows with no refitting. Where items were homogeneous, the replicate spread and the claimed uncertainty were close: the ratio of replicate spread to the median claimed standard error was 1.04 at the base generator and 1.09--1.13 at the three zero-heterogeneity settings, so the replicate spread exceeded the uncertainty the concept unit claims by 4--13\%. The agreement degraded under the concept scale effect in particular, where the ratio was 1.53 at $\omega = 0.5$ and 3.31 at $\omega = 1$. The ratio was not monotone in the threshold effect, at 1.39 at $\tau = 0.5$ but 1.12 at both $\tau = 1$ and $\tau = 2$, and it was 1.24 at skew $\alpha = 1$. In a typical dataset the interval then claimed less uncertainty than repeated runs showed, which is the anticonservative direction. Against the root mean square of the claimed standard errors, the average model standard error of Morris et al.~\cite{morris2019}, the ratio was 1.03 at $\omega = 1$ and 1.05 at $\omega = 2$. The difference between the two denominators reflects claimed standard errors that varied widely between datasets, with an extreme empirical tail: the largest 1\% of them contributed 54\% and 95\% of their mean square at $\omega = 1$ and 2, respectively. The near agreement of these second moments does not establish interval calibration or a generally adequate sampling unit. At $\tau = 0.5$ and $\alpha = 1$ the ratio was 1.36 and 1.23, respectively, against the root mean square as well. At $\omega = 2$ the tabulated median ratio of 71 should not be read as a measurement of that gap, because the replicate standard deviation there was dominated by rare extreme datasets, the same instability that makes the reference mean unusable at that setting. The family definitions suggest a source of this disagreement. In the audited families, M2S could absorb an atypical held-out concept scale, while the mixture members lacked that effect (Section~\ref{sec:example-families}). Under the Gaussian score, such a concept incurs a penalty for the selected mixture member that grows with the square of the scale ratio, so a single concept can dominate a dataset's joint log-score difference and push it in the graded direction (\SMomega). This mechanism follows from the member definitions and has not been isolated by a controlled simulation. The trial cannot be used as a finer unit. For any member with a concept effect, the held-out score of equation~\eqref{eq:q} is the joint likelihood of a concept's entire trial vector, so resampling trials independently is not a valid scheme for this statistic, and not merely too narrow a one (\SMaudit). The check does not identify heterogeneity as the cause of the interval's coverage shortfall, because the two diagnostics use the same rows and are not separated here.

\subsection{The Matched Comparison}\label{sec:results-matched}
The development stage ran the comparison on simulated graded participants before the design was frozen. With one noise covariance shared by both models, the two-state family won on data that were graded by construction, and six of 19 development sets read two-state. The estimated covariance understated the variance along the template direction by about 13\%. The repair estimates the covariance inside each model by maximum likelihood, as Section~\ref{sec:example-methods-matched} describes. Three further defects were found and repaired during development: starting points that missed the optimum by up to 22~nats, template coefficients that ran away along flat directions of the likelihood, and an expectation--maximization loop that stopped early (\SMjitter). Under the repaired fitter and the frozen rules, no development set of graded participants read two-state, and all 25 development sets of two-state participants (T1) read two-state. These readings are development evidence, and the calibration that follows used fresh seeds.

We then ran the frozen comparison on fresh simulated participants under all 12 instantiated settings (\SMjitter). T3 was absent: the search for an amplitude that reaches its target reference contrast found no bracketing pair of amplitudes. Under the four graded settings and four of the five departures, no set read two-state, and the reference contrast was negative in all eight, between $-0.00124$ and $-0.00107$~nat per trial. Under O5, 12 of 100 sets read two-state, and the reference contrast was $+0.00252$ [$+0.00203$, $+0.00301$]~nat per trial. Four W4 sets were unavailable because fewer than 16 of their 20 participants were usable. The graded-setting condition counts an unavailable set as a false positive, so that condition was not met. Under T1, 49 of 50 sets read two-state, with a one-sided 95\% lower bound of 0.909, so the detection condition was met. Under E1, no set fell below the margin, with a one-sided 95\% upper bound of 0.030, so the margin condition was met. The margin was $\epsilon = 0.01140$~nat per trial. E1 was certified at amplitude scale $\lambda = 1$. Certified means that the Monte Carlo interval of its re-estimated reference lay within $0.15\epsilon$ of the target (\SMjitter). The reference was $+0.01407$, with 97.5\% Monte Carlo interval [$+0.01286$, $+0.01528$]. The upper-bound assessment was conditional on this certification. All 50 T2 sets were unavailable, so the completion condition was not met. Hull coverage is the share of fits whose profile-likelihood hull for the latency spread contained the true spread (\SMjitter). At true latency spreads of 40 and 80~ms, hull coverage in the graded family was 0.848 and 0.802, respectively, and in the two-state family 0.792 and 0.726. All four rates fell below the 0.90 floor, and recovery of the latency spread was classified as not adequate in both families. The published per-window rule, applied to the same sets, called two states in 1 of 100 W1 sets, in 39 of 100 O5 sets and in 7 of 50 T1 sets, and in 0 of 100 sets under each of W3 and W4. Neither the W3 nor the W4 rate of two-state decisions rose above W1's under the declared comparison of two independent proportions, which calls a rise when the 95\% interval for their difference lies above zero (\SMjitter).

On the 20 participants of the auditory dataset, the held-out score difference between B3 and G3 was $+0.00457$~nat per trial, 95\% interval [$-0.00025$, $+0.00939$]. Without variation terms (B0 against G0) it was $+0.00421$ [$-0.00004$, $+0.00847$], and with the scale term only (B2 against G2) it was $+0.00608$ [$+0.00074$, $+0.01143$]. Subtracting the contrast with both variation terms from the contrast without either term gave $-0.00036$ [$-0.00396$, $+0.00324$]~nat per trial. Both contrasts used all 20 participants. The latency profile hull's lower bound exceeded 20~ms in no graded fit and one two-state fit, so neither family met the declared latency-detection condition. Because calibration did not meet the graded-setting or completion condition, the frozen rule reports these values descriptively. A separate reading, an upper bound, had its own conditions (an available margin, a certified E1 setting and the margin condition), and these were met. The upper end of the 95\% interval, $+0.00939$, lay below the margin $\epsilon = 0.01140$. The data therefore rule out a two-state advantage of B3 over G3 as large as the margin. This is an upper bound, not evidence that the two families fit equally well. The scale-only contrast and the other secondary contrasts make no decision of their own and are reported as values.

\subsection{The Bounded Comparison: Development Screens}\label{sec:results-bounded}
The candidate of Section~\ref{sec:example-methods-bounded} was screened on the saved rows before any new data were generated. The screen used the two families' existing equal-weight density ensembles, without the new scale member, and the bounded score at $\pm 0.03$~nat per trial. For each graded setting, 700 of the 1{,}000 saved datasets formed the calibration set and the remaining 300 the assessment set. In every setting, the interval contained the full-bank median of the bounded statistic in 299 or 300 of the 300 assessment datasets. All 3{,}600 assessment intervals lay below zero, and none called two states. Applied without changing the calibration, the development rule made two-state decisions in 199 of 200 saved datasets at the original gain of 0.01~nat and in 200 of 200 at 0.03~nat. At 0.003~nat, the counts were 49 of 50 for M3, 50 of 50 for M3L, and zero of 50 each for M3H and M3V. The changed comparison detected fewer two-state datasets at the weakest threshold-varying and unequal-variance alternatives, which include the case in which a varying threshold makes two states look graded~\cite{nieuwenhuis2011}. The new mixture member with a concept random scale and a richer start recipe were then checked locally on single development datasets (\SMbounded). These are point comparisons on single datasets and estimate no detection rate. The saved rows were used to choose the bound, so they cannot also validate it. A confirmatory decision still requires the calibration and validation on new seeds specified in Section~\ref{sec:example-methods-bounded}.

\subsection{A Development Run on an LLM}\label{sec:results-pilot}
We ran the procedure end to end on the LLM side, at the single layer of the false-positive study. We report that the procedure ran on real activations, and what it returned. It is a development run and is not evidence about access in LLMs, for the reasons given at the end of this subsection. Sixteen concepts were used to train the decoders, which were then frozen per layer. These concepts were chosen before the protocol assigned each concept of the bank to a role (\SMdesign{}). Sixteen further concepts supplied 2{,}688 held-out trials. They share no concept with the decoder-training concepts or with the sealed confirmatory set, which stays unused until the confirmatory study. Their readout passed the validation declared in advance. The adapter is the code that assembles the readouts into the dataset that the analysis function consumes. The checks of the capture (the recording of the residual stream at every layer), of the decoders and of the adapter are given in \SMpilot. The readouts were scored by the same analysis function as the 12{,}000 simulated datasets. At layer 41, inside the 35-layer workspace band (\SMdesign), with four inner starts and the fixed-score concept-cluster interval, every fit converged and no inner fit was non-finite. All three comparison variants returned graded:

\begin{center}
\small
\begin{tabular}{lrrr}
\toprule
\textbf{Comparison variant} & \textbf{Point} & \textbf{Interval} & \textbf{SE} \\
\midrule
Audited procedure (selected members) & $-0.1157$ & $[-0.1367,\, -0.0933]$ & 0.0111 \\
Equal-weight ensemble     & $-0.1212$ & $[-0.1366,\, -0.1048]$ & 0.0085 \\
Inherited pair            & $-0.0897$ & $[-0.1087,\, -0.0694]$ & 0.0103 \\
\bottomrule
\end{tabular}
\end{center}

\noindent All values are in nats per trial, the unit of equation~\eqref{eq:delta}, and SE is the claimed standard error of Procedure~\ref{proc:layer}. In each case, the mean of each of the eight concept families had the same sign as the mean over all concepts. The development run showed that the procedure ran on real activations with all five components specified and reported its score difference in nats per trial. The run's decision does not support an inference about access. The run used sixteen concepts of a development build of the stimuli, at one layer and in the active condition. The detection rate of Section~\ref{sec:results-detection} was shown on 64 concepts of simulated data, not on sixteen. The graded decision is therefore not read as evidence about access (\SMpilot). The run's score difference cannot be compared with the human plateau of Section~\ref{sec:results-human}, since the two readouts are different instruments on different systems. No confirmatory LLM data exist.

\section{Discussion}\label{sec:discussion}
Nothing in the cross-entropy audit depends on the substrate: the readout, the stimulus design and the five components are declared for each system in its own terms, and the same loop runs on both. Section~\ref{sec:discussion-general} discusses what the audit shows in general, Section~\ref{sec:discussion-example} what the example establishes, Section~\ref{sec:discussion-outcomes} what the held study's outcomes would mean, and Section~\ref{sec:discussion-limits} the limits of the completed analyses.

\subsection{What the Audit Shows in General}\label{sec:discussion-general}
The value of declaring the five components is easiest to see by asking what each result would be taken to mean without them. Read alone, the human result would be quoted as all-or-none access in the brain from about 250~ms, the article's own figure, or from 315~ms, the first window above 0.95 in its source data under the authors' score convention. The reproduction qualifies that sentence in four ways (Section~\ref{sec:results-human}): the effect moves with the fitting, depends on the rule and on the scale of the scores, and is numerically small. Without the components declared, it reads as a firm physiological fact. Read alone, the LLM-side observation is that an LLM's decoded share (Section~\ref{sec:example-adaptation}) sits near one endpoint or the other and switches abruptly at a threshold, with bimodal decoded shares across prompts~\cite{gurnee2026}. Under the inherited pair of models, that pattern is the kind of data a two-state model is fitted to. The false-positive study of Section~\ref{sec:results-false} is a direct caution about that inference. The inherited pair returned two states on datasets generated with a single graded state, and those decisions concentrated where the generator gives items different thresholds. Gurnee et al.\ align the mixing weight to each prompt's own threshold. Their prompts mix the embeddings of two country names. The threshold differs between country pairs, as it differs between concepts here. It also differs between carrier sentences within a pair, a variation that neither family here represents. The caution applies to their histogram pooled across prompts, and the width of each prompt's own transition is evidence within one item that our audit does not assess. Items in a stimulus bank generally differ from one another~\cite{rajalingham2018}, and a fixed checkpoint has no per-participant fitting to absorb the difference. The false-positive study does not evaluate any published measurement, none of which used this procedure. It establishes that data which the inherited models read as two-state need not come from two states. The inherited models read a single graded process with item heterogeneity as two-state.

The components matter most when the two decisions are compared, because the same decision on both systems is the result most likely to be over-read. The two decisions concern different populations: the human unit is the participant and the inference concerns new participants, the LLM-side unit is the concept and the inference concerns new concepts at one checkpoint. The variability they summarize has different origins, neural variation between trials in the brain and variation placed in the stimulus by design on the machine. Two systems returning the same decision would therefore share a statistical model of their readouts, not a mechanism~\cite{geirhos2020}, and would do so over different populations. That distinction is available only because the sampling unit and the estimand were declared. Without them, two systems that return the same decision may be read as systems of the same kind. The common unit, in turn, means that both results are stated on one scale and audited against the same criteria. The human preference is 0.0025--0.003~nat per trial, and an LLM-side value is reported on the same scale and not in activation units. The unit is common, but the effect it measures is not. The two applications use different readouts, designs and competing families, so they define different predictive tasks. We therefore do not compare the development run's magnitude with the human plateau. A graded decision can count against alternatives for which the procedure has demonstrated a high detection rate at the tested design size. The false-negative uncertainty of that rate limits how much the decision counts. The detection study establishes such a rate only for its twelve declared settings. The resemblance between the two literatures also becomes checkable. At present, the correspondence between all-or-none entry in the brain and abrupt switching in an LLM is an agreement of vocabulary. Whether it also holds in measured quantities can be tested by a measurement made on both systems through one declared interface. That interface is the readout, the stimulus design and the five components.

Two consequences follow for indicator claims in general. First, an error rate measured at the new system's design is a precondition for an indicator claim built on a transferred test, and the audit supplies it where the frameworks of Section~\ref{sec:related-construct} set their thresholds by argument. Second, the audit's output is evidence about the procedure. Construct validity, whether the decision indexes the property, still needs the evidence that the measurement literature asks for (Section~\ref{sec:problem-two}). A second audited test needs its own declared readout, sampling design and calibration. For the two indicators that fall inside the class, metacognitive sensitivity and the aggregate ignition index, \SMdeclarations{} gives the declarations an audit would start from, and agreement on such declarations makes the assumptions explicit before decisions are compared.

\subsection{What the Example Establishes}\label{sec:discussion-example}
On the brain side, the published competing-model test of the two-state prediction reproduced in our re-implementation on the authors' data on the three points of Section~\ref{sec:results-human}, with differences at single windows, among them those where the effect begins. The passive session, analysed here as an addition, ranked the graded model highest, but no window passed the Simes correction for any model. That ranking admits the alternative readings of Section~\ref{sec:discussion-limits} and does not demonstrate graded physiology. The matched comparison applied the audit to the human test itself (Section~\ref{sec:results-matched}). Its calibration on simulated participants was completed before the real-data contrast was read. It detected two-state data in 49 of 50 T1 sets, and no E1 set fell below the margin. Its development stage found and repaired a defect: with one noise covariance shared by both models, graded data read two-state in 6 of 19 sets. The calibration did not meet two of the conditions that a decision on the real contrast requires. Under O5, whose data-generating model was unknown (Section~\ref{sec:example-methods-matched}), 12 of 100 simulated sets read two-state where at most one was allowed. Four W4 sets were unavailable as well, so the graded-setting condition was not met. All 50 T2 sets were unavailable, so the completion condition was not met. The frozen rule therefore reports the real-data values descriptively. A separate upper-bound reading rests on its own conditions, which were met (Section~\ref{sec:results-matched}). It places the upper end of the 95\% interval for the two-state advantage on the real data below the calibrated margin. The published per-window rule called two states in 39 of the same 100 sets, against 12 for the matched comparison. Neither count establishes whether either comparison identified two-state structure or model misspecification. On the same sets, the literature's tests found no dip anywhere. They read two Gaussian components wherever a lognormal scale term or pooling across participants shaped the pooled distances. They found the variance-burst signature on the resampled recordings and the real data, and on at most 3 of 100 sets of any fitted generator (Section~\ref{sec:results-false}). Read afterwards on the two-state question, the matched comparison was the only one of these procedures that made no two-state decision under the graded settings and the departures O1 to O4 and detected 49 of 50 T1 sets.

On the LLM side, the audit establishes properties of the decision procedure, not of any LLM. The procedure made no false two-state decision under twelve declared graded settings and detected the declared mixtures on 599 of 600 datasets. At six settings its bias-eliminated coverage fell below the 0.90 floor (Section~\ref{sec:results-interval}). The interval therefore cannot be used for confirmation, and any replacement has to be validated on new seeds under the graded generators that most resemble a mixture. Development screens on saved simulation rows supported a bounded comparison as the candidate for that validation (Section~\ref{sec:results-bounded}). The methodological point has two parts. A check of error rates alone does not reveal this failure: the procedure made no false two-state decision at any of the settings where its coverage fell short. And what an interval targets must be stated and shown to be estimable. It is established that naive cross-validation intervals for one model's error miss their target~\cite{bates2024cv}. The need to name the estimand is also established~\cite{morris2019}. In brain-machine comparison, a standard interval for error consistency had to be replaced by simulation under a declared null~\cite{geirhos2020}. What is specific here is this procedure and this failure under declared graded generators that resemble a mixture. Also specific is the difficulty of estimating the interval's reference at $\omega = 2$, where rare extreme datasets dominate (Section~\ref{sec:results-interval}). We report no confirmatory result from LLM captures. The development run showed only that the procedure runs on real activations with all five components specified (Section~\ref{sec:results-pilot}). Our LLM-side results give no evidence on whether the LLM's representations in the workspace band are two-state or graded under a natural-text dose, and we infer no claim about phenomenal consciousness from these analyses.

\subsection{What the Held Study's Outcomes Would Mean}\label{sec:discussion-outcomes}
The LLM study is held for a separate report, and its outcomes are not read here. \SMdesign{} and \SMaccounts{} record the design and the additional evidence needed to interpret its possible outcomes. Two limits on any such reading are fixed already. Matching measurements on two implementations would support only the limited claim that the two systems behave alike on that measurement, not a law that holds on every substrate or an identity of mechanisms. Any link to phenomenal consciousness would rest on further assumptions that these measurements do not supply. The outcome mapping of \SMaccounts{} is our reading, and, following the adversarial collaboration's protocol~\cite{melloni2023}, the registered report that will carry the held study will seek the workspace authors' endorsement of what each outcome would mean before the report is frozen.

\subsection{Limits of the Completed Analyses}\label{sec:discussion-limits}
The limits are these. (i) The human reference used one dataset of twenty participants, one auditory protocol and one decoder, and its inherited per-window models described amplitude only. The matched comparison was built to answer the latency objection~\cite{truccolo2002} on the same recordings. Its calibration did not meet the graded-setting and completion conditions that a decision on the real data requires. (ii) The held-out scores carry the leakage of the published design, in which information from the scored trials enters the fitting. This reproduction inherits that leakage and does not quantify it. Starting values are computed from all trials, the decoder's and the likelihood's cross-validation are not nested, and the decoder folds of the released code are not block-disjoint. (iii) The two-state preference of the human reference was found only in the task session. Suppose that late signatures split into two states when participants report and vanish when they do not. The two-state preference would then be linked to report~\cite{cohen2024noreport}. A criterion applied to a graded signal predicts the same pattern. A passive threshold at the top of the level range~\cite{sergent2021}, or access without a late response under low task relevance~\cite{whyteSmith2021}, would instead make the passive ranking uninformative about access. The data here do not separate these readings. (iv) Each level pools two vowels and blocks use different noise files~\cite{sergent2021}, and trial thresholds vary with pre-stimulus state~\cite{vanVugt2018}. Within-participant threshold heterogeneity of the kind that produced 651 false positives at $\tau = 2$ is therefore present in the human design, and no human family represents it. (v) The simulation audit covered one layer, one design size and twelve graded generators, each a retained member of the family the procedure fits, with four starts for inner selection and eight for refitting. No interval of the audited procedure came within $0.0067$~nat per trial of the decision boundary. The four mixture generators carried no concept-level scale heterogeneity, which only the audited graded family could represent. (vi) On both substrates the readout is a decoder fitted at the extremes of the evidence and applied near threshold. Whether two-state structure survives that projection, or whether the projection can itself produce it, is not tested here. Classifier readouts carry a bias toward the trained contrast~\cite{kriegeskorte2008}, decoders lose late phases when transferred~\cite{marti2017}, rating instruments shift mass to their ends~\cite{overgaard2006,nieuwenhuis2011,windeyCleeremans2015}, and the human readout may blend graded early and late components~\cite{mashour2020,vanVugt2018}. Answering the question needs a simulation that generates activations and projects them, which our audit does not do. (vii) The model families omit four features of the theory's own description of the near-threshold system. The low state has no dose dependence in any mixture member, although stimuli that participants report as unseen evoke dose-graded activity~\cite{delCul2007,dehaeneNaccache2001} and their processing continues after the divergence from trials reported as seen~\cite{sergent2005}. The inherited two-state model allows no high state at the catch level, the level without signal. Spontaneous entries and false alarms would place a high state there~\cite{kouider2010}. No family lets the threshold vary from trial to trial, and no member has more than two components, although two states have proved too few for a global-state transition~\cite{proektHudson2018} and several independently accessed clues could give more states~\cite{kouider2010}. The first and fourth omissions, the dose-free low state and the two-component limit, bias the comparison toward graded decisions. The third, the trial threshold, could produce false two-state decisions. A two-state decision would also not identify the number of states. Generators of each kind belong in the frozen generator catalogue of the bounded comparison (Section~\ref{sec:example-methods-bounded}). (viii) The held LLM study has further hazards that belong to its report (\SMdesign): one quantized checkpoint (its weights stored at reduced numerical precision), a correlational and token-defined lens, a dose that is a count of clauses, and a layer axis that is not time. Properties of the stimuli could also produce a mixture with no two-state process in the LLM, and no generator in the audit carries such a stimulus property. (ix) The experiment reported from the largest adversarial test of the workspace theory registered no prediction about the near-threshold distribution~\cite{cogitate2025}, so neither a two-state nor a graded result would settle the dispute over the theory. A second open dataset, on near-threshold visual detection with seen and unseen reports~\cite{melcon2024,melcon2025ds006171}, is planned as an extension from hearing to vision. Its analysis protocol is in development, and no outcome has been computed.

\section{Conclusion}\label{sec:conclusion}
The aim of this paper was an instrument for a fair comparison of brain experiment data with the measurements of an artificial or any other system. The cross-entropy audit is the method proposed for it. Two procedures were audited: the matched comparison on the brain side and the audited procedure on the LLM side. Neither yet meets all of its own declared conditions. On the same simulated data, these two procedures combined two results that no other procedure we ran matched. They made no two-state decision on graded data, and they detected two-state data in 49 of 50 T1 sets and 599 of 600 LLM-side datasets (Sections~\ref{sec:results-false} and~\ref{sec:results-detection}). On the brain side, the matched comparison read two-state on no set of the four graded settings or of the departures O1 to O4, which challenge its families (Table~\ref{tab:brainside}). Each of the other procedures run on the same sets fell short on at least one of those two results. The dip test detected none of the 200 two-state sets. The comparison of Gaussian mixtures by BIC called two states on 90 of 100 graded sets that carry no trial-to-trial variation. The published study's variance burst appeared on none of the 50 fitted two-state sets, and the published per-window rule called two states on 7 of them. On the LLM side, the inherited pair, the human test's two models carried over unchanged, returned two states on 989 of 12{,}000 graded datasets. The comparison of Gaussian mixtures by BIC did so on 5{,}577 of them, and the audited procedure on none. The dip test made no two-state decision on graded data either, but it detected only 18 of the 600 two-state datasets, where the audited procedure detected 599. The one dataset the audited procedure did not detect came from the setting where failure would be expected first. That setting is the mixture member with a free catch-level occupancy, set to 0.05, at the weakest gain of 0.003~nat per trial. The procedure returned an inconclusive interval there, not a graded decision. The rules for the literature's tests were declared before the tests ran, but no ranking of the procedures was fixed. This comparison is therefore a reading of their declared counts, made afterwards. It considers the two-state question alone, which not every one of those tests was built to answer. The comparison shows why a transferred test's error rates have to be measured at the new system's design before its decision is read: each of the other procedures with a decision rule erred, in one direction or the other, on data whose answer was known. The audit makes that measurement possible because it scores every competing model of either system on one scale, the difference in held-out cross-entropy in nats per trial, and declares the five components through which a test reaches its decision. Nothing in it depends on the substrate. On the brain side, our re-implementation on the open data of twenty participants reproduced the published comparison on the three points this paper uses: the first crossing of 0.95 at 315~ms under the authors' score convention, the order of the three models across time and the 435--495~ms plateau. It differed only at single windows, which enter no result. The reproduced active-session preference was modest, 0.0025--0.003~nat per trial, and its boundaries in time moved with the fitting and with the scale of the scores. The procedure also ran end to end on real activations, as a development run. Three limits frame these counts. We applied the audit to one test, the two-state access test moved from human EEG to an LLM, and we do not show that it carries to another test of this kind. On the LLM side, the graded settings were members of the family the audited procedure fits, and its interval contained the mean of the estimates in fewer than 90\% of datasets at six of the twelve graded settings. That interval is the one component to be repaired before a confirmatory decision. On the brain side, the matched comparison read two-state on 12 of 100 sets resampled from the recorded trials, whose data-generating model is unknown. The per-window rule read 39 of them as two-state. Four W4 sets were also unavailable. The declared graded-setting condition covers both the resampled sets and the W4 sets, and it was not met. All 50 T2 sets were unavailable, so the completion condition was not met either. Because these two conditions were not met, the real contrast is reported descriptively. Its upper bound rests on separate conditions, which were met, and it places the upper end of the 95\% interval below the calibrated margin. None of these counts says that the published test is false, and they are not a false-positive rate in general. The next steps are to repair the interval and validate it on new seeds, to run the confirmatory comparison of the two systems on that repaired instrument, and to carry the audit to a second transferred test, metacognitive sensitivity or the aggregate ignition index.

The confirmatory comparison of the two systems, for which the example was built, is a further study. It begins with the held LLM study at one layer on its sealed stimuli, once the repaired interval has been calibrated and validated on new seeds. A later stage reads a band of 35 layers in place of one, after its own calibration. That study is a large computation: one component alone, a complete screen of the refitting interval, would cost about 470{,}000 core-hours at the measured per-resample rate (\SMrepair). It is in progress and will be reported separately, and the audit reported here was run so that the study will rest on an instrument whose error rates have been measured. Any further indicator built on a transferred test needs the same: its own declared readout, sampling design and calibration, and an error rate measured at the new system's design before its decision is trusted. For the two further indicators inside the class, metacognitive sensitivity and the aggregate ignition index, \SMdeclarations{} gives the declarations an audit would start from, and the audit loop has not been run on either. Declaring the components allows a procedure to be examined, criticized and repaired on its own terms before any system is judged by it. Human cognition is one well-studied biological case within a wider question: how physical systems process what they receive and make it available for use. Within that question, the audit scores the competing models of a brain and of a system of any other kind on one scale. That scale makes the two results reportable side by side. It does not make their effects comparable in size (Section~\ref{sec:audit-scale}). Matching measurements would show only that the two systems behave alike on that measurement, and any link to phenomenal consciousness would need assumptions these measurements do not supply (Section~\ref{sec:discussion-outcomes}).

\section*{Data and Code Availability}\label{sec:availability}
The EEG recordings and the authors' analysis scripts are the open release of Sergent et al.~\cite{sergent2021} on the Open Science Framework (project aw3t5). The re-implementation's code, the 12{,}000 false-positive-study rows, the detection study's rows and their analysis scripts are public in the recoverable-self-coding repository~\cite{rsc_code} and archived in version~2 of the study snapshot~\cite{study_snapshot_v2}. \SMarchive{} gives, for every reported analysis, its paths, commits, checksums and runtimes, and states which later results require an additional archived release. No confirmatory LLM captures exist. Generative AI tools based on LLMs, Claude Code (Anthropic) and Codex (OpenAI), were used under the author's direction to write and test the analysis and simulation code. They were also used to draft and revise the text, and to check its claims, equations and numbers against the code and the saved outputs. The tools are not authors. The author directed the work and takes full responsibility for the content.

\section*{Supplementary Materials}
The following supporting information can be downloaded with the article as one PDF, Supplementary Materials S1--S16: S1, Proposed LLM Study: Design; S2, Human Reproduction: Conventions, Diagnostics and the Passive Session; S3, Simulation Audit: Conventions and Provenance; S4, The Reference at $\omega = 2$; S5, The Detection Study: The Blinded Reading; S6, The Development Run: Capture, Decoder and Adapter Checks; S7, The Competing Accounts; S8, Related Approaches: The Comparison Table; S9, Information Quantities and Thermodynamics: What Is Not Claimed; S10, The Matched Comparison: Calibration Design; S11, The Refitting Interval: Stages, Amendments and Cost; S12, Archive Inventory of the Development Results; S13, Two Further Indicators: What an Audit Would Have to Declare; S14, The Bounded Comparison: Specification, Prerequisites and Development Checks; S15, Other Summaries of the Simulated Readouts; S16, Index of Symbols and Settings. Everything in S1 is proposed and has not been run, and no confirmatory LLM data exist.

\authorcontributions{Conceptualization, methodology, formal analysis, software, and writing: P.v.R.}
\funding{This research received no external funding.}
\institutionalreview{This work is a secondary analysis of a de-identified human EEG dataset released publicly by its original investigators. No new human data were collected, and no participant was recruited or contacted. The original collection was validated by the CERES ethics committee of Paris Descartes University, as reported in the source publication~\cite{sergent2021}. Ethical review and approval were waived for this secondary analysis: on 21 October 2026 the Stellenbosch University Health Research Ethics Office confirmed in writing that studies of data already in the public domain are exempt from ethics approval under the South African National Department of Health, NHREC and HREC guidelines.}
\informedconsent{Informed consent was obtained from all participants by the original investigators of the publicly released dataset analysed here~\cite{sergent2021}. No new participants were involved in this work.}
\dataavailability{The human EEG recordings and the original analysis scripts are the authors' open release~\cite{sergent2021}. The re-implementation's code, the 12{,}000 false-positive-study rows behind the audit and later results reported here are archived in version~2 of the study snapshot~\cite{study_snapshot_v2} (DOI \href{https://doi.org/10.5281/zenodo.23513442}{10.5281/zenodo.23513442}). Those results are the independent reference bank (the further 1{,}000 datasets at $\omega = 2$ of Section~\ref{sec:results-interval}), the post hoc analyses of the false-positive-study rows, the detection study, the development run on the LLM, the development interval and timing summaries, and the concept-bank eligibility count of \SMdesign{}. The snapshot's files are those of the public code repository~\cite{rsc_code} at tag \texttt{entropy-access-v2}. For most results it holds the saved rows, the analysis script and the saved output. The development run is archived as its outputs and provenance. Its inputs, the activation captures and the LLM checkpoint, are not redistributed, so it cannot be recomputed from the archive alone. \SMarchive{} gives the paths, checksums and runtimes for these archived analyses. The matched comparison of Sections~\ref{sec:example-methods-matched} and~\ref{sec:results-matched} and the same-data comparisons of Section~\ref{sec:results-false} postdate version~2 and are not included in that snapshot. Three post hoc analyses of Section~\ref{sec:results-human} also postdate it, namely the selection without the null model, the total held-out score and the passive-session description. The replicate-spread comparison against the root mean square of the claimed standard errors and the finite-sample resampling factor, both of Section~\ref{sec:results-interval}, the interval summaries of Sections~\ref{sec:results-false} and~\ref{sec:results-interval} and the bounded comparison's development screens of Section~\ref{sec:results-bounded} postdate it as well. Their code, design, generator assets and development, calibration and real-data records require an additional archived release.}
\conflictsofinterest{The author is an entrepreneur and inventor involved in the development of artificial intelligence and computational systems. These activities did not influence the analysis or conclusions presented in this work. The author declares no conflict of interest.}
\acknowledgments{The author thanks the investigators of the open EEG release analysed here for publishing their data and analysis code. During the preparation of this manuscript, the author used Claude Code (Anthropic) and Codex (OpenAI) to write and test the analysis and simulation code. The tools also assisted with drafting and revising the text, and with checking its claims, equations and numbers against the code and the saved outputs. The author has reviewed and edited the output and takes full responsibility for the content of this publication.}

\reftitle{References}
\bibliography{refs}

\end{document}